\documentclass[review]{elsarticle}

\usepackage{lineno,hyperref}

\journal{Journal of Computational Statistics and Data Analysis}

\usepackage{natbib}
\usepackage{amsmath}
\usepackage{graphicx,psfrag,epsf,setspace}
\usepackage{enumerate}

\usepackage{url} 

\usepackage{graphicx}
\usepackage{epsfig}
\usepackage{amsmath}
\usepackage{amssymb}
\usepackage{caption}
\usepackage{color}
\usepackage{enumerate}
\usepackage{multicol}
\usepackage{float}
\usepackage{subfig}
\usepackage{appendix}
\usepackage{setspace}

\newcommand{\eps}{\mbox{$\epsilon$}}
\newcommand{\be}{\begin{equation}}
\newcommand{\ee}{\end{equation}}
\newcommand{\ba}{\begin{eqnarray}}
\newcommand{\ea}{\end{eqnarray}}
\newcommand{\bpm}{\begin{pmatrix}}
	\newcommand{\epm}{\end{pmatrix}}

\newcommand{\beq}{\begin{equation}}
\newcommand{\eeq}{\end{equation}}
\newcommand{\bea}{\begin{eqnarray}}
\newcommand{\eea}{\end{eqnarray}}
\newcommand{\p}{^{\prime}}
\newcommand{\bs}{\boldsymbol}
\newcommand{\indep}{\rotatebox[origin=c]{90}{$\models$}}
\newcommand{\nt}{\notag} 

\newcommand{\bsbeta}{\bs{\beta}}
\newcommand{\bseta}{\bs{\eta}}
\newcommand{\bsX}{\bs{X}}
\newcommand{\bsy}{\bs{y}}

\newcommand{\kappasq}{\kappa^2}
\newcommand{\lambdasq}{\lambda^2}

\newcommand{\bseps}{\bs{\epsilon}}

\DeclareMathOperator*{\argmin}{arg\,min}

\begin{document}

\begin{frontmatter}

\title{Bayesian Analysis of Spatial Generalized Linear Mixed Models with Laplace Moving Average Random Fields}


\author{Adam Walder \corref{mycorrespondingauthor}}
\author{Ephraim M. Hanks} 
\cortext[mycorrespondingauthor]{Corresponding author}
\address{Pennsylvania State University, University Park}
\ead{arw39@psu.edu}

\begin{abstract}
		Gaussian random field (GRF) models are widely used in spatial statistics to capture spatially correlated error. We investigate the results of replacing Gaussian processes with Laplace moving averages (LMAs) in spatial generalized linear mixed models (SGLMMs). We demonstrate that LMAs offer improved predictive power when the data exhibits localized spikes in the response. SGLMMs with LMAs are shown to maintain analogous parameter inference and similar computing to Gaussian SGLMMs. We propose a novel discrete space LMA model for irregular lattices and construct conjugate samplers for LMAs with georeferenced and areal support. We provide a Bayesian analysis of SGLMMs with LMAs and GRFs over multiple data support and response types.  
\end{abstract}


\end{frontmatter}

	\section{Introduction}
	Gaussian processes are the most common method for describing spatially and/or temporally correlated errors. The Gaussian random field (GRF) possesses an intuitive dependence structure, as well as a globally flexible fit. These desirable features have popularized the use of Gaussian processes in spatial and temporal statistics, as well as design of experiments and other fields. Despite the GRF's flexible nature, Gaussian processes can over-smooth in the presence of local spikes \cite{paciorek2004nonstationary}. In this work we consider the use of Laplace moving average models (LMAs) in place of traditional Gaussian processes in spatial generalized linear mixed models (SGLMMs) 
	
	LMAs have received sporadic attention over the past decade as alternatives to GRFs \cite{aaberg2011class, wallin2015geostatistical,opitz2016modeling}. However, there has been no systematic comparison of LMAs and GRFs for spatial generalized linear mixed models (SGLMMs) in the Bayesian framework. Our contributions in this work include 
	\begin{enumerate}
		\item The development of a novel discrete space (areal) LMA model for irregular lattices.
		\item The construction of conjugate samplers for both continuous (point-referenced) and discrete (areal) SGLMMs with LMAs. 
		\item A Bayesian analysis comparing the predictive power and computational efficiency of LMAs and GRFs over a range of scenarios, including continuous, binary, and count data collected both in discrete (areal) and point-referenced (geostatistical) spatial support.  
	\end{enumerate} 
	
	Whittle \cite{whittle1954stationary} demonstrated that continuous space GRFs with Mat\'ern covariance arise as solutions to a stochastic partial differential equation (SPDE). Lindgren et al. \cite{lindgren2011explicit} constructed a sparse finite element representation of this Gaussian Mat\'ern SPDE. As a result, a sparse form of the multivariate normal distribution can be used to fit Mat\'ern GRF models in a computationally efficient manner \cite{rue2001fast}. Bolin \cite{bolin2014spatial} extended the finite element approximation of Lindgren et al. \cite{lindgren2011explicit} to the case of Type-G Mat\'ern random fields of which the LMA is a special case. In Section \ref{Section_Laplace_Random_Fields}, we provide a summary of this extension for the symmetric LMA. Following Bolin \cite{bolin2014spatial}, the LMA can similarly be expressed as a conditionally sparse Gaussian random field through the introduction of auxiliary data. We also provide insights for handling the computational issues associated with MCMC implementation. 
	
	Wallin and Bolin \cite{wallin2015geostatistical} explored the LMA for geostatistical data with Gaussian responses. Though the discrete space model was claimed to be analogous, no further exploration was considered. Faulkner and Minin \cite{faulkner2018locally} provided a Bayesian implementation of the graph trend filtering (GTF) estimates of Wang et al. \cite{wang2016trend} for temporal data. Both works, found that replacing traditional Gaussian priors for Laplace priors provided a model with better adaptivity in the presence of local ``spikes" in the response. We develop a novel discrete space analog to the continuous space LMA model that is an extension of Wang et al. \cite{wang2016trend} for SGLMMs. Our model can easily be implemented in place of any Gaussian conditionally autoregressive (CAR) or simultaneously autoregressive (SAR) model. In Section \ref{Section_Discrete_Space}, we propose a novel MCMC implementation of our Bayesian hierarchical model for discrete space SGLMMs.

	The LMA models of this paper offer an intuitive alternative to traditional GRF SGLMMs. The discrete space and continuous space LMA models are constructed based on sparse matrix operations making for fast and efficient fitting. We provide Bayesian analyses based on our novel MCMC implementation of the LMA models over four separate data sets. Our MCMC implementation and model construction allows for Bayesian inference with LMAs that is just as interpretable as with GRF models. In some cases, the LMA is shown to provide a better fit than the Gaussian model. Given the ease of implementation, and familiarity of inference, LMA models can be useful alternatives to GRF models.  
	
	The paper is organized as follows: In Section 2 we provide background material needed to develop our models. In Section \ref{Section_Laplace_Random_Fields} we discuss finite element approximations for continuous space LMAs. We also provide details related to the numerical issues involved with fitting LMAs via MCMC. In Section \ref{Section_Discrete_Space} we detail our discrete space LMA model and its relation to the GTF estimates of Wang et al. \cite{wang2016trend}. In Section \ref{Section_Example_Analysis} we consider four datasets on which the LMA model is compared to its GRF counterpart. We conclude with a discussion in Section \ref{Section_Discussion}. 
	
	
	\section{Spatial Models with Mat\'ern Random Fields} 
	In this section we provide background information to assist in developing the framework of the hierarchical spatial models considered in this work. We begin by describing the SGLMM, and follow with a discussion of the Mat\'ern random field as a solution to a stochastic partial differential equation (SPDE). 
	
	\subsection{SGLMMs} \label{Section_SGLMM}
	Generalized linear models (GLMs) model the mean $\bs{\mu}$ of a distribution $f$ with linear predictors through an invertible link function $g(\cdot)$. Generalized linear mixed models (GLMMs) are GLMs that allow for the inclusion of random effects in the linear predictor. The spatial GLMM (SGLMM) attempts to capture an unobserved spatially varying trend by imposing a dependence structure in the random effect. 
	
	Let $\{\bs{u}_i\}_{i=1}^{N}$ be a collection of locations observed in some spatial domain $\Omega \subset \mathbb{R}^d$. Consider $Y(\bs{u}_i)  \sim f(y_i)$ such that the mean, $\mu_i = E\left(Y(\bs{u}_i)\mid\bsbeta, \eta(\bs{u}_i), \eps(\bs{u}_i) \right)$, is modeled through the link function 
	\bea
	g(\mu_i) = \bs{x}_i\p \bsbeta + \eta(\bs{u}_i) + \eps(\bs{u}_i). \label{SGLMM_Link} 
	\eea 
	In some cases an uncorrelated random effect, $\eps(\bs{u})|\sigma^2 \sim \mathcal{N}(0,\sigma^2)$, is included. In the Gaussian response case, $\sigma^2$, is thought of as homogeneous measurement error. The spatially varying random effect $\eta(\bs{u}) | \bs{\theta}$ is included in \eqref{SGLMM_Link} to capture spatial dependence. The hyper-parameters ($\bs{\theta}$) govern the mean and covariance structure of the spatial random effect. The most common distribution assumed for $\eta(\bs{u}) | \bs{\theta}$ is Gaussian. This work considers replacing the traditional Gaussian prior for a less common Laplace moving average (LMA). We provide a thorough comparison of the two prior choices over four datasets in Section \ref{Section_Example_Analysis}.

	\subsection{Random Fields with Mat\'ern Covariance} \label{Subsection_Random_Fields_with_Matern_Covariance}
	Stationary Mat\'ern random fields arise as stationary solutions to the SPDE
	\bea 
		(\kappasq - \triangle)^{\alpha/2} \eta(\bs{u}) = \xi \mathcal{W}(\bs{u}), \quad \bs{u} \in \mathbb{R}^d, \quad \alpha = \nu + d/2 \label{SPDE_GRF_eq}
	\eea 
	where $\triangle = \sum_{i=1}^{d} \frac{\partial^2}{\partial u_i^2}$ is the Laplacian operator, $\kappa$ is a spatial scale parameter, $\alpha$ controls the smoothness of the realized fields, and $\xi$ is a variance parameter. The SPDE in \eqref{SPDE_GRF_eq} is driven by Gaussian white noise, $\mathcal{W}(\bs{u})$. Whittle \cite{whittle1954stationary,whittle1963prediction} showed that stationary solutions to \eqref{SPDE_GRF_eq} are GRFs with Mat\'ern covariance
	\bea 
		C(\bs{u},\bs{v}) = \frac{\phi^2}{2^{\nu-1}\Gamma(\nu)} (\kappa ||\bs{v}-\bs{u} ||)^{\nu} K_{\nu}(\kappa ||\bs{v}-\bs{u} ||), \quad \bs{u}, \bs{v} \in \mathbb{R}^d \label{Matern_cov}
	\eea 
	where $||\cdot||$ denotes the Euclidean norm in $\mathbb{R}^d$. The marginal variance is $\phi^2 = \left( \xi^2 \Gamma(\nu) \right) / \left(\Gamma{(\alpha)}(4\pi)^{d/2}\kappa^{2\nu}\right)$ and the approximation of the effective range is given by $\rho = \sqrt{8\nu}/\kappa$ \cite{lindgren2011explicit}. Lindgren et al. \cite{lindgren2011explicit} provided a finite element representation of the SPDE in \eqref{SPDE_GRF_eq}. In Section \ref{Subsection_Finite_Element_Approximations_to_Gaussian_Matern_Random_Fields} we detail Lindgren's approximation. The result is a sparse Gauss Markov random field (GMRF) approximation to a GRF with Mat\'ern covariance given in \eqref{Matern_cov}. 
	
	\section{LMA Models for Point Referenced Data} \label{Section_Laplace_Random_Fields} 
	
\begin{figure}[H]
		\centering
		\includegraphics[scale=0.5]{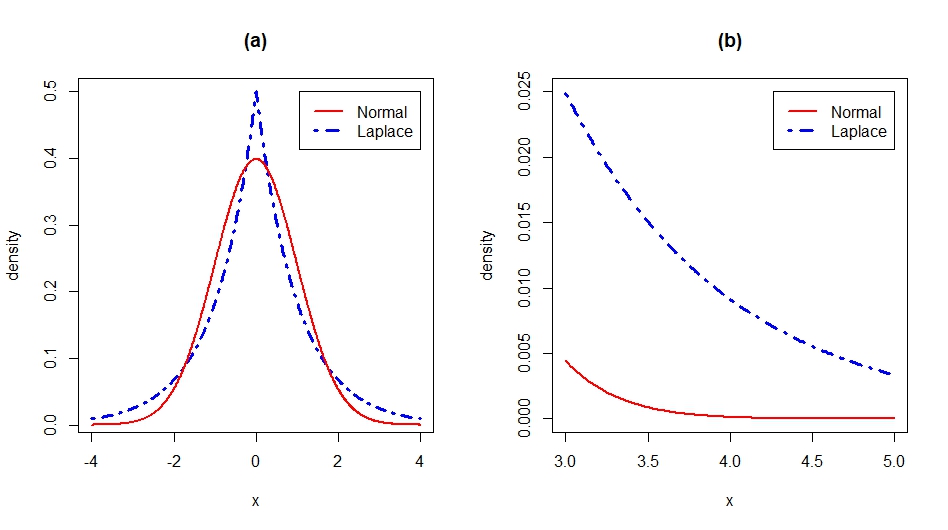}
		\caption{(a) Standard normal, $\mathcal{N}(0,1)$, and scale one Laplace density plots. (b) Tails of the respective distributions.}
		\label{Laplce_vs_Normal_Dist}
	\end{figure}

	GRF modeling remains appealing to scientists due to familiarity of implementation and the intuitive dependence structure of Gaussian processes. The Laplace distribution offers a wider tailed, sharper peaked, alternative to the Gaussian distribution (see Figure \ref{Laplce_vs_Normal_Dist}). In this Section we provide an overview of the results of Bolin \cite{bolin2014spatial}, which demonstrate that the LMA can be expressed as the solution to an SPDE similar to its Gaussian counterpart. We then lay out the finite element approximations of Lindgren et al. \cite{lindgren2011explicit} and Bolin \cite{bolin2014spatial} which provide sparse representations of the analytic solutions of SPDEs driven by Gaussian and Laplace noise respectively. We conclude the Section by providing a novel exploration of model fitting via MCMC. 
	
	\subsection{Laplace Moving Average Models as SPDEs} \label{Subsection_LMAs_as_SPDEs}
	Gaussian priors often produce marginal distributions with light tails. \.Aberg and Podg\'orski \cite{aaberg2011class} suggested the use of LMAs to obtain asymmetric and heavier tailed marginals. \.Aberg and Podg\'orski \cite{aaberg2011class} showed that the LMA can be expressed as a convolution of a kernal with Laplace noise. Bolin \cite{bolin2014spatial} showed that LMA models with Mat\'ern covariance can equivalently be expressed as the solution to an SPDE by replacing the Gaussian white noise, $\mathcal{W}(\bs{u})$, in \eqref{SPDE_GRF_eq} with Laplace noise. Laplace noise on a compact set $\Omega \subset \mathbb{R}^d$ can be expressed as 
	\bea 
		\dot{\Lambda} = \sum_{k=1}^{\infty} \left( \Gamma_k + \sqrt{\Gamma_k} G_k \right) \delta_{\bs{u}_k}  \label{Eqn_Laplace_Noise} 
	\eea 
	where $G_k \stackrel{iid}{\sim} \mathcal{N}(0,1)$, $\Gamma_k \sim e^{-\nu \gamma_k } W_k$, $W_k$ are \textit{iid} standard exponentials, $\gamma_k$ are arrival times of a Poisson process with intensity one, and $\delta_{\bs{u}_k}$ is a Dirac distribution centered at $\bs{u}_k$ with each $\bs{u}_k$ uniformly distributed in $\Omega$ (see Bolin \cite{bolin2014spatial} for details).  
	
	The SPDE defining an LMA process, $\eta(\bs{u})$, with Mat\'ern covariance is given by
	\bea 
		(\kappasq - \triangle)^{\alpha/2} \eta(\bs{u}) =  \dot{\Lambda}(\bs{u}), \quad \alpha = \nu + d/2. \label{SPDE_LRF_eq}
	\eea

	Both the SPDEs driven by Gaussian and Laplace noise produce random fields with Mat\'ern covariance. However, the SPDE driven by Laplace noise provides a random field with the ability to capture ``spikey" spatial behavior better than GRFs. Unlike its Gaussian counterpart, there does not exist a closed form solution for the Laplace driven SPDE in \eqref{SPDE_LRF_eq}. Bolin \cite{bolin2014spatial} used a finite element approximation of the SPDE in \eqref{SPDE_LRF_eq} to fit LMA models using an EM algorithm. In Sections (\ref{Subsection_Finite_Element_Approximations_to_Gaussian_Matern_Random_Fields}--\ref{Subsection_Finite_Element_Approximations_to_Matern_LMAs}) we detail the finite element approximations to the Gaussian and LMA SPDEs proposed by Lindgren et al. \cite{lindgren2011explicit} and Bolin \cite{bolin2014spatial}  respectively.

	\subsection{Finite Element Approximations to Mat\'ern GRFs}  \label{Subsection_Finite_Element_Approximations_to_Gaussian_Matern_Random_Fields}
	The finite element representation of the LMA model proposed by Bolin \cite{bolin2014spatial} is an extension of the results in the Gaussian case. For the sake of comparison, we first detail the finite element approximation of Lindgren et al. \cite{lindgren2011explicit}.
	
	In Section \ref{Subsection_Random_Fields_with_Matern_Covariance} we expressed GRFs with stationary Mat\'ern covariances as analytic solutions to SPDEs. Though the analytic solution provides useful insights, model fitting and parameter estimation are often facilitated by considering a numerical approximation. Lindgren et al. \cite{lindgren2011explicit} proposed the use of a finite element approximation to the stochastic weak formulation of the SPDE 
	\bea 
		(\kappasq - \triangle)^{\alpha / 2}\eta(\psi) = \xi \mathcal{W}(\psi), \quad \alpha = \nu + d/2 \label{Eqn_SPDE_Test_Gauss}
	\eea 
	where $\{\psi\}$ is a set of test functions. The finite element method (FEM) solution begins by expressing the solution, $\eta(\bs{u})$, as a basis expansion 
	\bea 
		\eta(\bs{u}) = \sum_{i=1}^{n} \phi_i(\bs{u}) w_i, \quad \bs{u} \in \Omega \label{basis_expansion}
	\eea  
	where $\{\phi_i(\bs{u})\}_{i=1}^{n}$ is a set of basis functions on $\Omega$. The solution in \eqref{Eqn_SPDE_Test_Gauss} is only required to hold for a finite collection of $\psi_i$. The Galerkin method approximate solution is obtained by setting $\{\psi_i\}_{i=1}^{n} = \{\phi_i\}_{i=1}^{n}$. 
	
	Lindgren et al. \cite{lindgren2011explicit} formulated an FEM approximation by considering $\{\phi_i(\bs{u})\}_{i=1}^{n}$ to be piecewise triangular basis functions.  The basis functions are constructed by partitioning the spatial region of interest, $\Omega \subset \mathbb{R}^d$, into non-overlapping triangular regions. The corners of the triangles, referred to as vertices, are assigned $n$ Gaussian weights, denoted $w_i$. Each $\phi_i$ is defined to be 1 at vertex \textit{i} and 0 at all other vertices. Lindgren et al. \cite{lindgren2011explicit} derived the distribution of the weights 
	\bea 
		\bs{w} | \xi,\kappa \sim \mathcal{N}\left( \bs{0}, \xi^2 \bs{Q}^{-2}_{\alpha} \right) \label{weight_dist} 
	\eea 
	where $\bs{Q}_{\alpha}$ is determined by the choice of $\alpha$ in \eqref{Eqn_SPDE_Test_Gauss}.
	
	Following Lindgren et al. \cite{lindgren2011explicit}, the precision matrix in \eqref{weight_dist} is defined as  
	\begin{equation}
		\bs{Q}_{\alpha} = \begin{cases} 
		\bs{Q}_{1} = \bs{L}, & \alpha = 1  \\
		\bs{Q}_{2} = \bs{L}\bs{C}^{-1}\bs{L}, &  \alpha = 2  \\
		\bs{Q}_{\alpha} = \bs{L}\bs{C}^{-1}\bs{Q}_{(\alpha-2)}\bs{C}^{-1}\bs{L}, &  \alpha \geq 3  
		\end{cases}
	\label{cont_Q}
	\end{equation}
	where $\bs{L}=\kappasq \bs{C} + \bs{G}$. The matrices used to define \textbf{L}, are given by 
	\bea 
		C_{ij} &=& \int_{\Omega} \phi_i(\bs{u})\phi_j(\bs{u}) d\bs{u} \label{C_not_Sparse}\\
		G_{ij} &=& \int_{\Omega} \nabla \phi_i(\bs{u}) \nabla \phi_j(\bs{u}) d\bs{u}. \label{G_Mats}
	\eea 
	
	Note that \textbf{C} as defined in \eqref{C_not_Sparse} is sparse, but its inverse, $\bs{C}^{-1}$, which is required in $\bs{Q}_{\alpha}$ for $\alpha \geq 2$, may not be. In turn, the resulting precision matrix, $\bs{Q}_{\alpha}$, may not be sparse. To ensure sparsity in $\bs{Q}_{\alpha}$, Lindgren et al. \cite{lindgren2011explicit} provided a GMRF approximation to the GRF representing the numerical solution to the SPDE in \eqref{Eqn_SPDE_Test_Gauss} by replacing \textbf{C} in \eqref{C_not_Sparse} with the diagonal matrix,
	\bea 
		C_{ii} = \int_{\Omega} \phi_i(\bs{u}) d\bs{u}	\label{C_sparse}
	\eea 
	
	Under lattice refinement, the sparse representation of \textbf{C} as defined in \eqref{C_sparse} converges to the same solution given by \textbf{C} in \eqref{C_not_Sparse} (see Lindgren et al. \cite{lindgren2011explicit} Appendix C.5). \textbf{C} in \eqref{C_sparse} is now a diagonal matrix relating to the volume of the regions produced by the mesh. \textbf{G} is a sparse matrix with zero entries on the diagonal, describing the connectivity of the mesh nodes.
	
	\subsection{Finite Element Approximations to Mat\'ern LMAs} \label{Subsection_Finite_Element_Approximations_to_Matern_LMAs}
	
	Bolin \cite{bolin2014spatial} extended the results of Lindgren et al. \cite{lindgren2011explicit} to the case of Type-G Mat\'ern random fields. We consider the FEM approximation for the symmetric LMA model with Mat\'ern covariance. Similar to the Gaussian case, the FEM approximation begins with a stochastic weak formulation of the SPDE in \eqref{SPDE_LRF_eq} given by 
	\bea 
		(\kappasq - \triangle)^{\alpha/2} \eta(\psi) = \lambda \dot{\Lambda}(\psi) \label{Eqn_Weak_LMA_SPDE}
	\eea 
	Following Section \ref{Subsection_Finite_Element_Approximations_to_Gaussian_Matern_Random_Fields} we construct piecewise linear basis functions $\{\phi_i\}_{i=1}^{n}$. Let $\Gamma_i \sim Gamma\left(\tau C_{ii}, \lambdasq\right)$ where $C_{ii}$ is the $i^{th}$ element of \textbf{C} in \eqref{C_sparse}. Define $\bs{\Gamma} = diag(\Gamma_1,...,\Gamma_n)$. Bolin \cite{bolin2014spatial} showed that the distribution of the basis expansion weights given by the Galerkin method is 
	\bea 
		\bs{K}_{\alpha}\bs{w} | \bs{\Gamma} &\sim& \mathcal{N}(\bs{0},\bs{\Gamma} ) \label{weights_prior_LRF_cnt}
	\eea 
	where
	\begin{equation}
		\bs{K}_{\alpha} = \begin{cases} 
		\bs{K}_{2} = \bs{L}, & \alpha = 2  \\
		\bs{K}_{\alpha} = \bs{L}\bs{C}^{-1}\bs{K}_{\alpha-2}, &  \alpha = 4,6,8,...   \\ 
		\end{cases}
		\label{K_alpha}
	\end{equation}
	with $\bs{L} = \kappa^2 \bs{C} + \bs{G}$ defined as in the Gaussian case. We note that this Section contains all the information needed to construct and fit an FEM solution to LMA models. We refer the reader to Bolin \cite{bolin2014spatial} for a mathematically rigorous construction of the FEM solution. 
	
	There currently exists no extension of the LMA approximation for odd valued $\alpha$'s. We also point out that \eqref{weights_prior_LRF_cnt} results in a sparse precision matrix for small values of $\alpha$. For $\alpha = 2$, the precision matrix, $\bs{K}_2 \bs{\Gamma}^{-1} \bs{K}_2$, defines a sparse representation for the roughest covariance function offered by the finite element approximation to the LMA. This corresponds to Mat\'ern covariance with $\nu = 1$ for spatial models in $\mathbb{R}^2$ and $\nu = 1.5$ for $\mathbb{R}$.
	
	In summary, we see that the LMA model can be expressed as a sparse conditionally Gaussian distribution conditioned on Gamma-distributed auxiliary variables. We discuss the numerical issues and computational considerations implied by this approximation in Section \ref{Subsection_Model_Fitting_Cnt_Space}.
	
	\subsection{Model Fitting: Continuous Space} \label{Subsection_Model_Fitting_Cnt_Space}
	Parameter estimation of LMAs is difficult since no closed form of the likelihood exists. \.Aberg and Podg\'orski \cite{podgorski2011estimation} proposed a method of moments-based estimation for LMA models. In Section \ref{Subsection_Finite_Element_Approximations_to_Matern_LMAs} we showed that expressing the LMA as an SPDE allows for inference in the likelihood framework. Bolin \cite{bolin2014spatial} constructed an EM algorithm for parameter estimation following the FEM approximation described in Section \ref{Subsection_Finite_Element_Approximations_to_Matern_LMAs}. Wallin and Bolin \cite{wallin2015geostatistical} considered the use of an MCEM algorithm in order to provide a computationally efficient extension to SGLMMs. Persistent numerical issues contributed to difficult parameter estimation, in both the works of Bolin \cite{bolin2014spatial} and Wallin and Bolin \cite{wallin2015geostatistical}. The issues stem from the fact that $\text{E}\left[\Gamma_i^{-1}|\bs{ \cdot} \right]$ is unbounded for small $\underset{1 \leq i\leq n}{\min} |\tau C_{ii} - 1/2 |$. Bolin \cite{bolin2014spatial} suggested truncating the expected value at 1000 to avoid numerical instabilities.     
	
	To our knowledge, there has been no systematic Bayesian comparison of SGLMMs with LMAs and GRFs fit via MCMC. Samplers for GRFs require samples from the full-conditionals of the $n$-dimensional basis expansion weights $\bs{w}$, variance parameter $\xi$, spatial scale parameter $\kappasq$, fixed effects $\bs{\beta}$, and a homogeneous random effect variance $\sigma^2$ (if applicable). Samplers for the continuous space LMA require \textit{n} additional auxiliary variables ($\Gamma_i's)$ and a shape parameter $\tau$. 
	
	For the sake of illustration, consider a continuous response point referenced model with \textit{N} unique observed locations denoted $\{\bs{u}_i\}_{i=1}^{N}$. Define the \textit{N} by \textit{n} projection matrix $(\bs{A})$ with entries $A_{ij} = \phi_j(\bs{u}_i)$, where $\{\phi_l(\bs{u})\}_{l=1}^{n}$ are triangular basis functions formed as described in Section \ref{Subsection_Finite_Element_Approximations_to_Gaussian_Matern_Random_Fields}. Using the basis expansion of $\eta(\bs{u})$ (see \eqref{basis_expansion} Section \ref{Subsection_Finite_Element_Approximations_to_Gaussian_Matern_Random_Fields}) and assuming homogeneous error measurement $\eps(\bs{u}_i) \stackrel{iid}{\sim} \mathcal{N}(0,\sigma^2)$, we can write the discretized likelihood as $[\bs{y}|\bsbeta,\bs{w},\sigma^2] \sim \mathcal{N}\left( \bs{X}\bsbeta + \bs{A}\bs{w}, \sigma^2 \bs{I} \right)$.   
	
	Conditioned on \textit{n} auxiliary gamma random variables, $\Gamma_i$, we express the LMA as a scale mixture of normals with gamma variance. For $\alpha =2$ it was shown in Section \eqref{Subsection_Finite_Element_Approximations_to_Matern_LMAs} equation \eqref{weights_prior_LRF_cnt} that $\bs{K}_2\bs{w}|\bs{\Gamma} \sim \mathcal{N}(\bs{0},\bs{\Gamma})$, where $\bs{\Gamma} = diag(\Gamma_1,...,\Gamma_n)$. The conjugate full-conditional for the weights of the finite element approximation of the LMA model are
	\bea 
		\left[\bs{w}|\bsy,\sigma^2,\bsbeta,\bs{\Gamma} \right] \sim \mathcal{N}\left( \left[ \sigma^2 \bs{L}\bs{\Gamma}^{-1}\bs{L} + \bs{A}\p \bs{A} \right]^{-1} \left[ \bsy - \bs{A}\bs{w} \right], \left[  \bs{L}\bs{\Gamma}^{-1}\bs{L} + \left(\frac{1}{\sigma^2}\right) \bs{A}\p \bs{A} \right]^{-1} \right). \label{Eqn_Gaussian_w_full_cond}
	\eea 
	
	Let $\bs{t} = \bs{K}_{\alpha} \bs{w} | \bs{\Gamma} \sim N(\bs{0},\bs{\Gamma})$ as defined in equation \eqref{weights_prior_LRF_cnt} of Section \ref{Subsection_Finite_Element_Approximations_to_Matern_LMAs}. This leads to conjugate generalized inverse Gaussian (\textit{GIG}) full conditionals for each gamma random variable, $\Gamma_i | t_i, \tau,\lambda \sim GIG\left(\tau C_{ii} -1/2, 2/\lambdasq, t_i^2\right)$, where the \textit{GIG(p,a,b)} density is given by 
	\bea 
	f(x;p,a,b) \propto x^{p-1} \exp\left(-\frac{ax + b/x}{2}\right). 
	\eea    
	Thus conjugate updates are available for the FEM weights $\bs{w}$ and the auxiliary variables $\Gamma_i$ in the LMA model. 
	
	In practice we found that Gibbs updates for the conjugate $\Gamma_i's$ resulted in poor-mixing (see Sections \ref{Data_Analysis_Malaria}--\ref{Data_Analysis_LAGOS}). This is not an uncommon issue, as conjugacy does not always produce an efficient sampler \cite{schliep2015data}. We found that the overall mixing of the Markov chains were improved by using independent one-at-a-time adaptively tuned Metropolis Hastings updates for each $\Gamma_i$.

	It is possible that samples from $\Gamma_i | t_i, \tau,\lambda$ are close to zero. This results in a numerically negative-definite precision matrix $\left[  \bs{L}\bs{\Gamma}^{-1}\bs{L} + \left(\frac{1}{\sigma^2}\right) \bs{A}\p \bs{A} \right]^{-1}$ for the full-conditional distribution in \eqref{Eqn_Gaussian_w_full_cond}. To overcome this issue, we re-sampled the $\Gamma_i's$ if $\left[  \bs{L}\bs{\Gamma}^{-1}\bs{L} + \left(\frac{1}{\sigma^2}\right) \bs{A}\p \bs{A} \right]^{-1}$ was numerically rank-deficient. This amounts to a Metropolis Hastings update for $\bs{w}$ and $\bs{\Gamma}$ with the constraint that $\left[  \bs{L}\bs{\Gamma}^{-1}\bs{L} + \left(\frac{1}{\sigma^2}\right) \bs{A}\p \bs{A} \right]^{-1}$ is positive-definite. 
	
	In binary error response SGLMMs, Gaussian full-conditionals can be constructed using data augmentation \cite{albert1993bayesian}. The LMA model can then be fit via MCMC exactly as described above. For Poisson error response distributions, the full-conditionals of $\bs{w}$ are not Gaussian. In this case, we suggest the use of conditionally independent one-at-a-time Metropolis Hastings updates for each $w_i | \bs{w}_{-i}$ (see Appendix \ref{Appendix_Conditionally_Independent_Block_Proposals}). We note that one-at-a-time conditionally independent block sampling is applicable for Gaussian and binary responses as well. However, in practice we found that block proposing $\bs{w}$ and $\bs{\Gamma}$ produced a faster and more efficient sampler for Gaussian and binary error responses. With the above numerical considerations we were able to fit the LMA models with a sampler that is a simple extension of traditional Gaussian samplers for each of point-referenced data analyses considered in Sections (\ref{Data_Analysis_Malaria}--\ref{Data_Analysis_LAGOS}).
	
	In summary, we have shown how LMA models with Mat\'ern covariance can be expressed as an SPDE. We demonstrated that the FEM approximations for the GRF and LMA models result in sparse conditionally Gaussian representations. Following the numerical cautions detailed in this Section, we were able to fit the LMA using MCMC. This allows for Bayesian analyses familiar to traditional SGLMM models with GRFs. 
	
	
	\section{LMA Models in Discrete Space} \label{Section_Discrete_Space} 
	
	In the previous Section, we detailed the construction of the LMA model and provided a method for parameter estimation via MCMC. Wallin and Bolin \cite{wallin2015geostatistical} acknowledged the potential use for LMAs in discrete space, but no further investigation was pursued. In this Section, we present a novel Bayesian hierarchical formulation for discrete space SGLMMs with LMAs. Our model is shown to be an adaptation of the widely recognized graph trend filtering (GTF) estimates proposed by Wang et al. \cite{wang2016trend}. Similar to the continuous space model, we introduce auxiliary variables to express the discrete space LMA as a conditionally Gaussian distribution. The resulting model exhibits computing and inference similar to its Gaussian analogue. 
	
	\subsection{Graph Trend Filtering} \label{Subsection_GTF}
	Heavier tailed alternatives to GRFs have been shown to provide improved predictive power \cite{tibshirani2014adaptive,faulkner2018locally,wang2016trend} for discrete space models. Wang et al. \cite{wang2016trend} extended the univariate trend filter of Kim et al. \cite{kim2009ell_1} to irregular graphs. We provide a Bayesian extension of the graph trend filer (GTF) that is analogous to the LMA model for discrete space SGLMMs. 
	
	To motivate our discrete space model, we provide a brief overview of GTF estimates. Let \textit{G}=(\textit{V},\textit{E}) be a graph with vertices $i = \{1,...,n\}$ and undirected edges $\{e_1,...,e_m\}$. Suppose that $\bs{y} = (y_1,...,y_n)$ are observed at the vertices. The GTF estimates, $\hat{\bs{\beta}} = (\hat{\beta}_1,...,\hat{\beta}_n)$, are the solution to 
	\bea 
		\hat{\bs{\beta}} = \argmin_{\bs{\beta} \in \mathbb{R}^n} \frac{1}{2} || \bs{y} - \bs{\beta} ||_2^2 + \lambda|| \triangle^{(k+1)}\bs{\beta} ||_1 \label{GTF_GTF_argmin}
	\eea  
	where $\triangle^{(k+1)}$ is a recursive graph difference operator of order \textit{k}, and $\lambda$ is a shrinkage parameter determined by cross-validation. For the case of $k=0$, $\triangle^{(1)}$ is defined such that $||\triangle^{(1)}\bs{\beta}||_1 = \sum_{\substack{ (i,j) \in E} } | \beta_i - \beta_j |$ produces a first order difference penalty over \textit{G}. The $k^{th}$ order GTF differencing matrix seen in \eqref{GTF_GTF_argmin} is recursively defined as
	\begin{equation}
		\triangle^{(k+1)} =
		\begin{cases}
		\left(\triangle^{(1)}\right)\p \triangle^{(k)}, & \text{for odd } k \\
		\triangle^{(1)} \triangle^{(k)}, &\text{for even } k
		\end{cases}. \label{GTF_triangle}
	\end{equation} 
	Higher orders of $k$ in \eqref{GTF_triangle} correspond to higher order differencing penalties. For a chain graph, or a graph over a one-dimensional line, the GTF estimates reduce to the trend filter estimates of Kim et al. \cite{kim2009ell_1}. 
	
	In this Section \ref{Subsection_GRFs_and_LRFs_in_Discrete_Space}, we propose a discrete space version of the LMA for SGLMMs. The GTF estimates are shown to be a special case of our model. In turn, our model specification serves as a generalized extension of the GTF to SGLMMs in the Bayesian framework.
	
	\subsection{LMAs in Discrete Space} \label{Subsection_GRFs_and_LRFs_in_Discrete_Space}
	Areal datasets are composed of aggregated responses. Examples of common spatial aggregations include cumulative measurements for cities, states, or countries. The areal units at which responses were recorded determine the discretization of space. The spatial discretization can be summarized by a graph with nodes at each areal unit and undirected edges defined by the spatial neighborhood structure of the areal units. 
	
	The graph detailing the spatial neighborhood relationships can be expressed as a graph Laplacian matrix, \textbf{A}, of the form 
	\begin{equation} A_{ij} = \begin{cases} 
		-1, & \text{\textit{i} is neighboring \textit{j}}\\
		\sum_{k} |A_{ik}|, & i = j \\
		0, & \text{else} 
		\end{cases}. \label{Graph_Lap}
	\end{equation}
	
	The graph Laplacian (\textbf{A}) serves as the precision matrix of the popular ICAR model for areal spatial random effect models \cite{besag1974spatial}. \textbf{A} is positive semi definite with rank \textit{n}-1, where \textit{n} is the number of observations. It is common to add a positive value to the diagonal to ensure that \textbf{A} is diagonally dominant. Define $\bs{L} = \kappasq \bs{I} + \bs{A}$. Let \textbf{D} be an upper triangular matrix such that $\bs{L} = \bs{D}\p \bs{D}$ (i.e., \textbf{D} could be a Cholesky decomposition). We consider the discrete space precision matrices given by, 
	\bea 
		\bs{Q}_k = \left( \triangle^{(k)} \right)\p \triangle^{(k)} \label{discrete_Q}
	\eea 
	where
	\begin{equation} 
		\triangle^{(k)} = \begin{cases} 
		\bs{L}^{\frac{(k+1)}{2}}, & \text{for odd \textit{k}} \\
		\bs{D}\bs{L}^{\frac{k}{2}}, & \text{for even \textit{k} } 
		\end{cases} \label{traingle_operator} 
	\end{equation} 
	
	Traditional simultaneously autoregressive (SAR) models (see Appendix \ref{Appendix_SAR}) assume a Gaussian prior for the \textit{n}-dimensional random effect. We assume a Gaussian prior on the weighted sum of differences  
	\begin{equation}
		\triangle^{(k)}\bseta | \xi^2,\kappasq \sim \mathcal{N}\left( 0,\xi^2 \bs{I} \right) \label{Eqn_Discrete_GRF_weight_prior} 
	\end{equation}
	which implies that 
	\begin{equation}
		Cov\left( \bseta|\xi^2,\kappasq \right) = \xi^2 \bs{L}^{-(k+1)} = \xi^2 \bs{Q}^{-1}_{k} \label{Eqn_Discrete_GRF_weight_cov} 
	\end{equation}
	
	The discrete space LMA is obtained by placing an \textit{iid} Laplace prior on each sum of weighted differences  $\triangle^{(k)}\bseta|\lambda,\kappasq$. That is,
	\begin{equation}
		{\triangle_l\p}^{(k)} \bseta | \lambda,\kappasq \stackrel{iid}{\sim} \mathcal{L}(\lambda), \quad l = 1,...,n \label{Eqn_Discrete_LRF_weight_prior}
	\end{equation}
	which implies that 
	\begin{equation}
		Cov\left(\bseta |  \lambda,\kappasq \right) = \frac{2}{\lambda} \bs{L}^{-(k+1)} = \frac{2}{\lambda} \bs{Q}_k^{-1} \label{Eqn_Discrete_LRF_weight_cov} 
	\end{equation}
	We observe that the discrete space LMA model is obtained by replacing the Gaussian prior in \eqref{Eqn_Discrete_GRF_weight_cov} for an \textit{iid} Laplace prior in \eqref{Eqn_Discrete_LRF_weight_cov}.
	
	\subsection{L2 vs. L1 Penalization} 
	In the proceeding Section we demonstrated that the discrete space LMA is obtained by replacing the Gaussian prior on the weighted sum of differences in \eqref{Eqn_Discrete_GRF_weight_prior} with \textit{iid} Laplace priors in \eqref{Eqn_Discrete_LRF_weight_prior}. Here we detail the penalization implications that result from altering the prior. The contrast in penalizations is presented to provide intuition as to why the LMA model should be considered in place of the the GRF model.
	
	Let $N(i) := \{j: j \text{ is a neighbor of } i\}$ and $\bs{\theta}_G = (\xi^2,\kappasq,\sigma^2)$. The order $k=1$ discrete space GRF has log full-conditional distribution	
	\bea 
		\log[\bseta|\bsbeta,\bsy,\bs{\theta}_G]	\approx \log[\bsy|\bsbeta,\bseta,\bs{\theta}_G] - \frac{1}{2 \xi^2}\sum_{i=1}^{n} (\kappasq \eta_i + \sum_{j \in N(i)}(\eta_i-\eta_j) )^2 + Const. \label{GRF_discrete_penalty}
	\eea
	Notice that \eqref{GRF_discrete_penalty}  resembles penalized regression with a squared penalty term placed on the weighted sum of differenced nodes ($\eta_i$). If $\kappasq > 0$ in \eqref{GRF_discrete_penalty}, $\bseta$ is penalized based on the sum of the localized differences relative to the magnitude of each node. We recover the intrinsic conditionally auto-regressive model (ICAR) by taking $\kappasq=1$ and $k=0$.
	
	Now define $\bs{\theta}_L = (\lambda,\kappasq,\sigma^2)$. The order $k=1$ discrete space LMA log full-conditional is given by 
	\bea 
		\log[\bseta|\bsbeta,\bsy,\bs{\theta}_L]	\approx \log[\bsy|\bsbeta,\bseta,\bs{\theta}_L] - \frac{\lambda}{2} \sum_{i=1}^{n} |\kappasq \eta_i + \sum_{j \in N(i)}(\eta_i-\eta_j)| + Const. \label{LRF_discrete_penalty}	
	\eea  
	
	 Our inclusion of $\kappasq$ allows for a full rank precision matrix, as well as an \textit{L}-1 penalty based on the magnitude of each node, $\eta_i$. The Bayesian equivalent to the $k^{th}$ order GTF estimates proposed by Wang et al. \cite{wang2016trend} are obtained by setting $\kappasq = 0$ in \eqref{LRF_discrete_penalty}. We also point out that for $\kappasq = 1$ and $k=0$, we recover the Laplace analog to the ICAR model.
	
	We note that the LMA model resembles a LASSO style (\textit{L}-1) penalty while the GRF model resembles a ridge like (\textit{L}-2) penalty on the sum of differenced nodes ($\eta_i$). There appears to be an analogous extension to be made for penalization in the continuous space. For a discrete space model on a regular grid, lattice refinement results in convergence to the continuous space LMA.

	\subsection{Model Fitting: Discrete Space} \label{Subsection_Model_Fitting_Discrete_Space}
	We aim to preserve the familiarity of the Gaussian fit for our discrete space model. To do so, we first recognize the Laplace distribution as a scale mixture of normals. This can be seen by taking a Gaussian variable, $Z \sim N(0,S_i)$, where $S_i$ is an independent exponential random variable with rate $\lambdasq / 2$. Marginalizing over $S_i$, it follows that $Z|\lambda \sim \mathcal{L}(\lambda)$. 
	
	Recall, the discrete space LMA places an $\mathcal{L}(\lambda)$ prior on the $l^{th}$ weighted difference, $\triangle^{(k)}_l \bseta | \lambda,\kappasq \stackrel{iid}{\sim} \mathcal{L}(\lambda)$. By introducing $n$ auxiliary variables $S_i \stackrel{iid}{\sim} Exp(\lambdasq/2)$, we can express the prior as  
	\bea 
		\triangle^{(k)} \bseta | \bs{S},\kappasq \sim N(0,\bs{S}), \quad \bs{S} = diag(S_1,..,S_n) \label{Eqn_Bayesian_GTF_prior_as_Normal}
	\eea 
	Park and Casella \cite{park2008bayesian} showed that $S_i^{-1}$ are conjugate inverse Gaussian ($InvGauss$). To see this, first define $\bs{t} = \triangle^{(k)} \bseta | \bs{S},\kappasq$. Then $S_i^{-1}|t_i,\lambda \sim InvGauss\left(\lambdasq,\sqrt{\frac{\lambdasq}{t_i^2}} \right)$, where the $InvGauss(a,b)$ has density
	\bea 
		f(x;a,b) = \left(\frac{b}{2 \pi x^3}\right)^{1/2} \exp\left(- \frac{b(x-a)^2}{2b^2x}\right) \label{Eqn_InvGauss_Density}
	\eea 
	
	We note that the \textit{InvGauss} is a special case of the \textit{GIG} with $p=-1/2$. We fit the discrete space LMA and GRF via MCMC as well. From \eqref{Eqn_Bayesian_GTF_prior_as_Normal}, we observe that the discrete space LMA model can be expressed as a conditionally sparse Gaussian distribution. Additionally, the full-conditionals for the auxiliary random variables are conjugate. In turn, Gibbs updates can be used for the full-conditionals of the \textit{n} auxiliary random variables. In Section \ref{Subsection_Model_Fitting_Cnt_Space} we discussed the numerical issues induced by using conjugate updates for the auxiliary mixing variables. This is not an issue in the discrete space sampler, as the \textit{InvGauss} does not require estimation of a shape parameter.  
	
	In summary, we have provided a novel discrete space Bayesian hierarchical LMA. The LMA is conditionally Gaussian, as in continuous space, again allowing for inference and computing familiar to GRF models. In Section \ref{Section_Example_Analysis} we demonstrate that the discrete space LMA offers improved out-of-sample predictive power in the presence of localized trends.

	
	\section{Example Analyses} \label{Section_Example_Analysis}
	In this section we compare the LMA and GRF models over four datasets; including Gaussian, Poisson, and binary responses observed both on continuous (point referenced) and discrete (areal) support. We perform 10-fold cross validation by randomly splitting the data set into 10 roughly equal sized groups. We withhold a test set $\bs{y}_k$ and train the model on the remaining observations $\bs{y}_{-k}$. We use the Bayesian cross validation scoring criterion (BCVS) of Hooten and Hobbs \cite{hooten2015guide} given by
	\bea 
		BCVS = -\sum_{k=1}^{10} \log\left( \frac{\sum_{t=1}^{T}[\bs{y}_k | \bs{y}_{-k},\bs{\theta}^{(t)}]}{T} \right)	
	\eea 
	where $T$ is the total number of stored MCMC iterations, and $\bs{\theta}^{(t)}$ is the $t^{th}$ sample of the model's parameters. Note that a smaller value implies a better model fit. 
	
	We used effective samples per second (ESS) to compare the computational performance of each model fit. All MCMC iterations had a burn-in phase in which the normal proposals were adaptively tuned according to Roberts and Rosenthal \cite{roberts2009examples}. Every data analysis involved 50,000 post-burn-in MCMC states. The number of burn-in iterations for the GRF and LMA model were held constant for each individual data analysis. The ESS was computed based on the likelihood evaluated at the stored 50,000 states. 
	
	\subsection{Discrete Space Examples: Slovenia Stomach Cancer} \label{Data_Analysis_Slovenia}
	We compare models trained on areal count data with the Slovenian stomach cancer outbreak dataset. The dataset consists of 194 responses of aggregated stomach cancer counts in each municipality of Slovenia collected from 1994 to 2001. The model proposed by Hodges and Reich \cite{hodges2010adding} was fit to investigate the relationship between standardized socioeconomic status, $SEc_i$, and the occurrence of stomach cancer, $y_i$. 
	\begin{figure}[H]
		\centering
		\includegraphics[scale=.8]{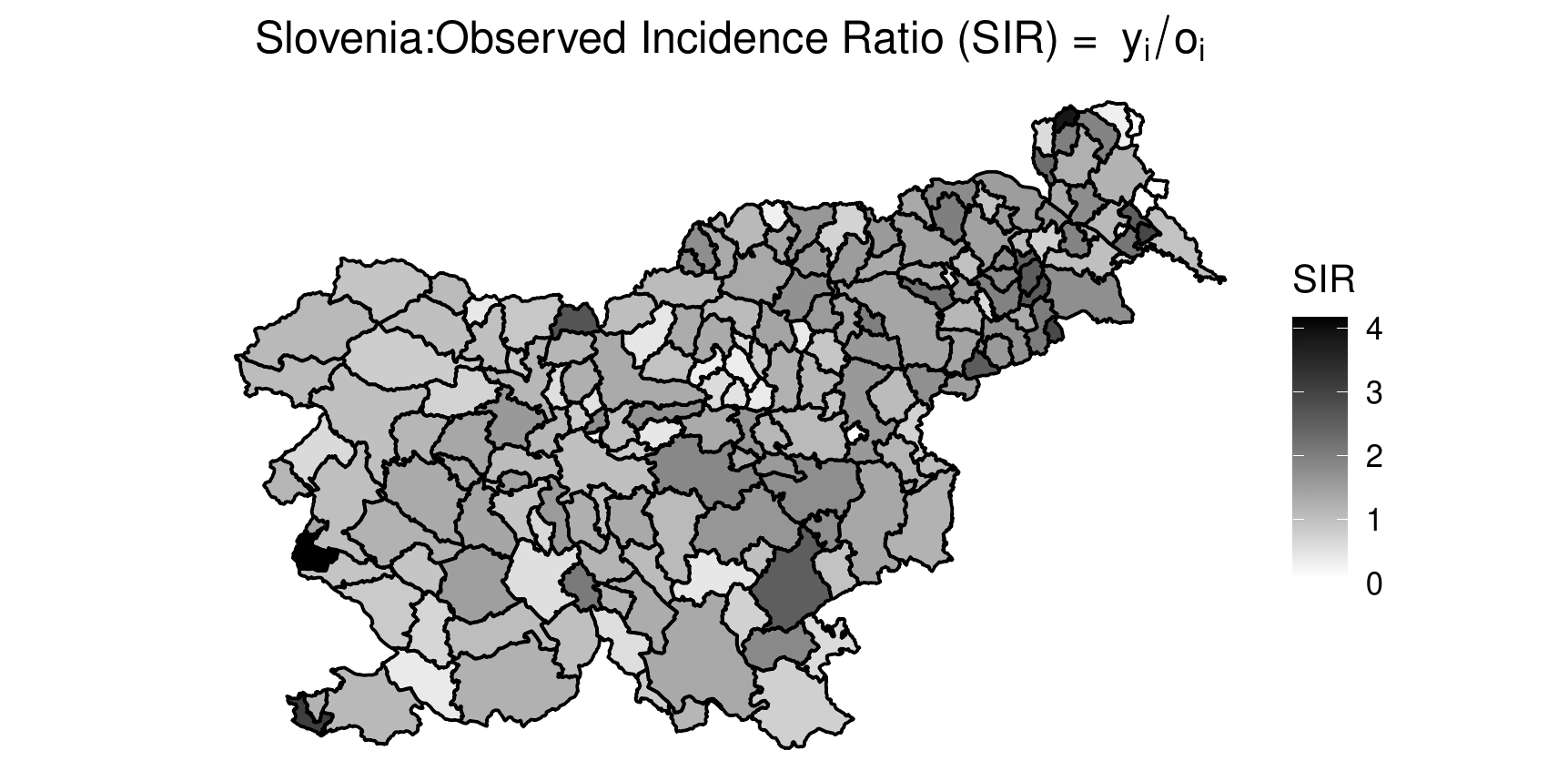}
		\caption{Plot of observed incidence ratio of stomach cancer (SIR), reported as the ratio of observed occurrences divided by the expected count in municipality \textit{i}.}
		\label{Slovenia_Plot}
	\end{figure} 
	We fit the Poisson spatial model considered by Hodges and Reich \cite{hodges2010adding} assuming stomach cancer counts are Poisson, $y_i \sim Poisson(\mu_i)$. The mean $\mu_i$ is  modeled through the log-link function  
	\bea 
		\log(\mu_i)= \log(o_i) + \bs{x}_i\p \bsbeta + \eta_i + \eps_i \label{Slovenia_Model}
	\eea 
	The offset term $(o_i)$ in \eqref{Slovenia_Model}, is the expected number of stomach cancer counts for observation \textit{i}. The covariate matrix $(\bs{X})$ contains an intercept and the standardized socioeconomic status ( SEc ). 
	
	We use an order $k=1$ differencing matrix $\triangle^{(1)}$ (see equation \eqref{traingle_operator} in Section \ref{Subsection_GRFs_and_LRFs_in_Discrete_Space}) to define the priors for the spatially correlated random effects in the GRF and LMA models defined in equations \eqref{Eqn_Discrete_GRF_weight_prior} and \eqref{Eqn_Discrete_LRF_weight_prior} of Section \ref{Subsection_GRFs_and_LRFs_in_Discrete_Space}. The fixed effects $(\bs{\beta})$ are assigned \textit{iid} normal priors with variance $10^6$. The variance of the spatially homogeneous random effect $(\sigma^2)$ is assigned an inverse gamma prior with shape and scale of one. The scale parameter for the GRF model $(\xi)$, the scale parameter for the LMA model $(\lambda)$, and the spatial scale parameter $(\kappa)$ are all assigned independent half-normal priors with scale one. The full-conditionals for the Poisson response data are detailed in Appendix \ref{Appendix_Poisson_Data}.
	
	Inference on the fixed effects is similar between the LMA and GRF models (see Table \ref{Discrete_Analysis_Table}). Table \ref{Slovenia_Table} shows that the GRF provides a better fit than the LMA. This is perhaps due to the areal dataset being smooth. We do note that the LMA sampler produces roughly the same ESS.   
	
	\begin{table}[H] \centering 
		\begin{tabular}{@{\extracolsep{5pt}} |ccc|} 
			\hline 
			Model & BCVS & ESS \\
			\hline
			GRF & $\textbf{560.61}$ & \textbf{4.12} \\ 
			LMA & $563.04$ & 3.81 \\
			\hline  
		\end{tabular} 
		\caption{BCVS and ESS for ten-fold cross validation on the Slovenia stomach cancer outbreak dataset.} 
		\label{Slovenia_Table} 
	\end{table}
	
	\subsection{Discrete Space Examples: Columbus Crime Data} \label{Data_Analysis_Columbus}
\begin{figure}[H]
		\centering
		\includegraphics[scale=1]{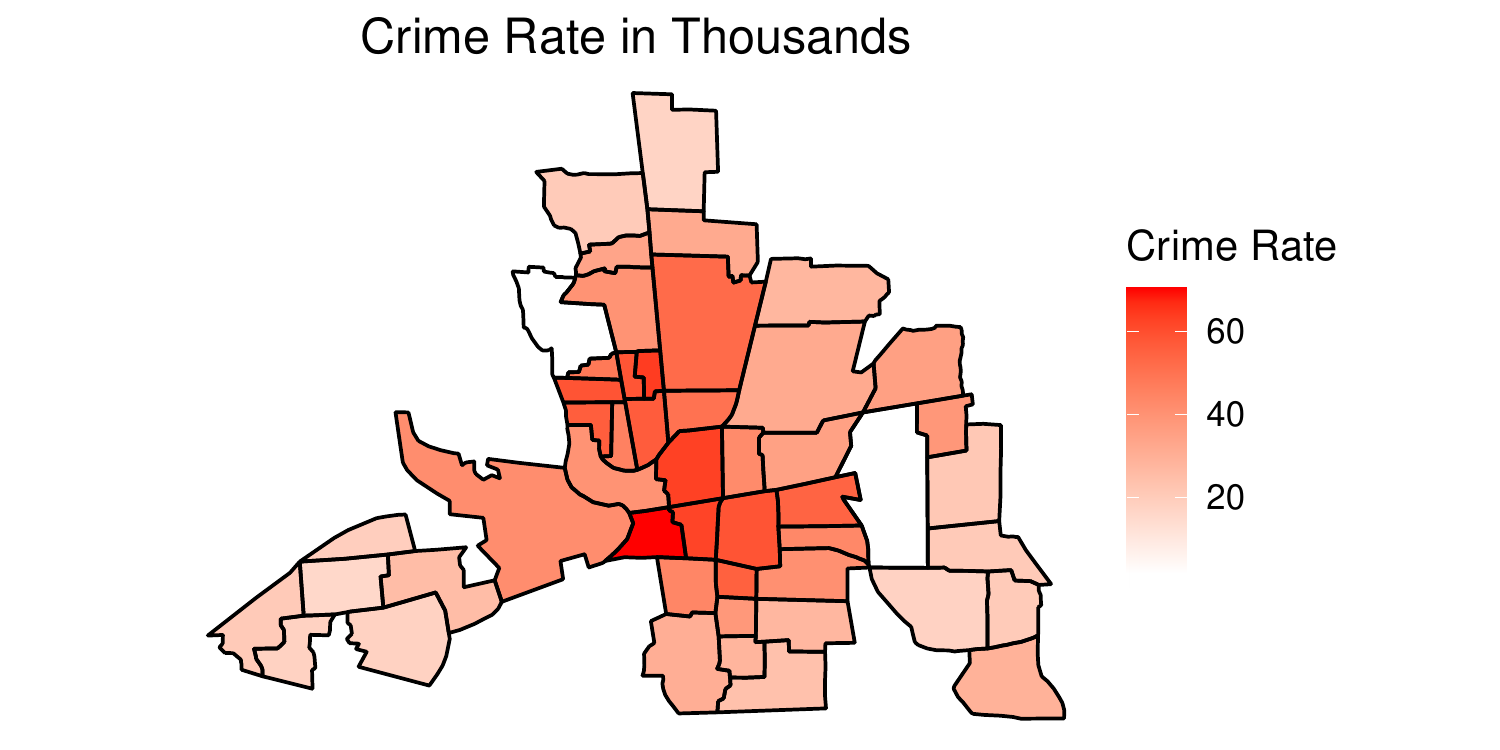}
		\caption{Plot of crime rate in thousands in the 49 counties of Columbus, Ohio.}
		\label{Columbus_Plot}
	\end{figure}

	The Columbus crime data are found in the \textit{``spdep''} R package \cite{spdep}. The dataset provides a spatial map with crime rates ($y_i$) for each county of Ohio. Ver Hoef et al. \cite{ver2018relationship} suggested modeling the data as an intercept only model with Gaussian response. We include average household income and average household value, as well as an intercept as covariates. Figure \ref{Columbus_Plot} shows a plot of the crime rates in each county of Ohio. There appears to be a few localized spikes in the data, namely the counties that appear in white. This is sufficient reason to suspect that the LMA model should provide an improved model fit. 
	The model is of form 
	\bea 
		y_i = \bs{x}_i\p\bsbeta +  \eta_i + \eps_i \label{Columbus_Dataset}  
	\eea 	
	We use the adjacency matrix (\textbf{A}) provided from the ``\textit{spdep}'' package to construct the first order differencing matrix $\triangle^{(1)}$ (see equation \eqref{traingle_operator} in Section \ref{Subsection_GRFs_and_LRFs_in_Discrete_Space}). The precision matrices for the priors of the GRF and LMA models are constructed from $\triangle^{(1)}$ as in Section \ref{Subsection_GRFs_and_LRFs_in_Discrete_Space}. We assumed \textit{iid} normal variance $10^3$ priors for the fixed effects. For both models $\kappa^2$ was assigned a half-normal scale one prior. The variance parameter for the GRF model $(\xi^2)$ and the variance parameter for the LMA model $(\lambda^2)$ were assigned half normal scale 10 priors. The full-conditionals for all GRF and LMA model parameters can be found in Appendix \ref{Appendix_Columbus_Crime_Data}.   
	
	In Table \ref{Columbus_Table}, we see that the LMA model is the clear favorite in terms of BCVS. Inference on the fixed effects $\bsbeta$ is similar in both cases (see Table \ref{Discrete_Analysis_Table}). The ESS for the LMA model seems substantially smaller than the GRF model. We note that the data set is only of size 49, so this difference may be a bit inflated. The ESS for both models is quite large for an MCMC sampler.   
	
	\begin{table}[H] \centering 
		\begin{tabular}{@{\extracolsep{5pt}} |ccc|} 
			\hline 
			Model & BCVS & ESS \\
			\hline
			GRF & $341.58$ & \textbf{18.13} \\ 
			LRF & $\bs{329.85}$ & 14.07 \\ 
			\hline
		\end{tabular} 
		\caption{BCVS from ten-fold cross validation and ESS for Columbus Ohio Crime dataset.} 
		\label{Columbus_Table} 
	\end{table} 
	
	\begin{table}[H]\centering \small  
		\scalebox{0.8}{ \begin{tabular}{@{\extracolsep{5pt}} cccccc}
			\hline \\[-1.8ex]
			\multicolumn{6}{l}{\textbf{Slovenia Stomach Outbreak}} \\
			\hline \\[-1.8ex] 
			Predictor & Parameter & GRF Estimate & 95\% CI & LMA Estimate & 95\% CI \\  
			\hline \\[-1.8ex] 
			Intercept & $\beta_0$ & $0.097$ & $(-0.036 , 0.202)$ & $0.115$ & $(-0.010, 0.339)$ \\ 
			SEc & $\beta_1$ & $-0.078$ & $(-0.170 , 0.031)$ & $-0.068$ & $(-0.162,0.040)$ \\ 
			\hline \\[-1.8ex] 
			& $\xi$ & $0.898$ & $(0.492, 1.561)$ & NA & NA \\ 
			& $\sigma$ & $0.290$ & $(0.238, 0.349)$ & $0.282$ & $(0.230, 0.339)$ \\
			& $\kappa$ & 1.427 & (0.616, 2.523) & 1.608 & (0.814, 2.543) \\ 
			& $\lambda$ & NA & NA & 1.127 & (0.718, 1.701)   \\
			\hline \\[-1.8ex]
			\multicolumn{6}{l}{\textbf{Columbus Crime Data}} \\
			\hline \\[-1.8ex]
			Predictor & Parameter & GRF Estimate & 95\% CI & LMA Estimate & 95\% CI \\  
			\hline \\[-1.8ex]
			Intercept & $\beta_0$ & 35.107 & (-27.661, 76.053) & 29.379 & (-39.597, 83.936) \\ 
			Avg. Inc &$\beta_1$ & -0.321  & (-0.391, -0.252) & -0.235 & (-0.327, -0.149) \\
			Avg. Value & $\beta_2$ & -0.981  & (-1.247, -0.716) & -1.057 & (-1.351, -0.755) \\
			\hline \\[-1.8ex] 
			& $\xi$ & 4.125 & (3.536, 4.676)  & NA & NA \\ 
			& $\sigma$ & 3.370 & (3.191, 3.550) & 2.803 & (2.541, 3.050)  \\
			& $\kappa$ & 0.162 & (0.051, 0.298)  & 0.219 & (0.081, 0.373)  \\ 
			& $\lambda$ & NA & NA & 4.582 & (4.086, 5.040)    
		\end{tabular} }
		\caption{Parameter estimates for discrete space data analysis examples of Sections (\ref{Data_Analysis_Slovenia}--\ref{Data_Analysis_Columbus})}
		\label{Discrete_Analysis_Table} 
	\end{table} 
	
	\subsection{Continuous Space Examples: Malaria in the Gambia, Africa} \label{Data_Analysis_Malaria}
	A model comparison for the binary continuous case is illustrated with the use of presence absence data of malaria in the Gambia, Africa. The data was made publicly available by Diggle and Riberio \cite{diggle2007model} in the \textit{``geoR"} package of \textit{R} \cite{rgeos}. The dataset consists of 2035 children records recorded at 65 village locations denoted $\{\bs{u}_i\}_{i=1}^{65}$. Each village, \textit{i}, has $n_i$, respondents.  We consider a model similar to that of Hanks et al. \cite{hanks2015restricted}. Let $y^{(i)}_j$ be the indicator for the presence ($y^{(i)}_j=1$) of malaria in the $j^{th}$ child at village location $i$. The covariates considered for each child are composed of the child's age, an indicator of whether or not a bed net was used, an indicator for whether or not an insecticide was applied to the bed net, the log normalized difference vegetation index (NDVI) at each village, and an indicator of presence or absence of a health center in the village. 
	
\begin{figure}[H]
		\centering
		\includegraphics[scale=0.75]{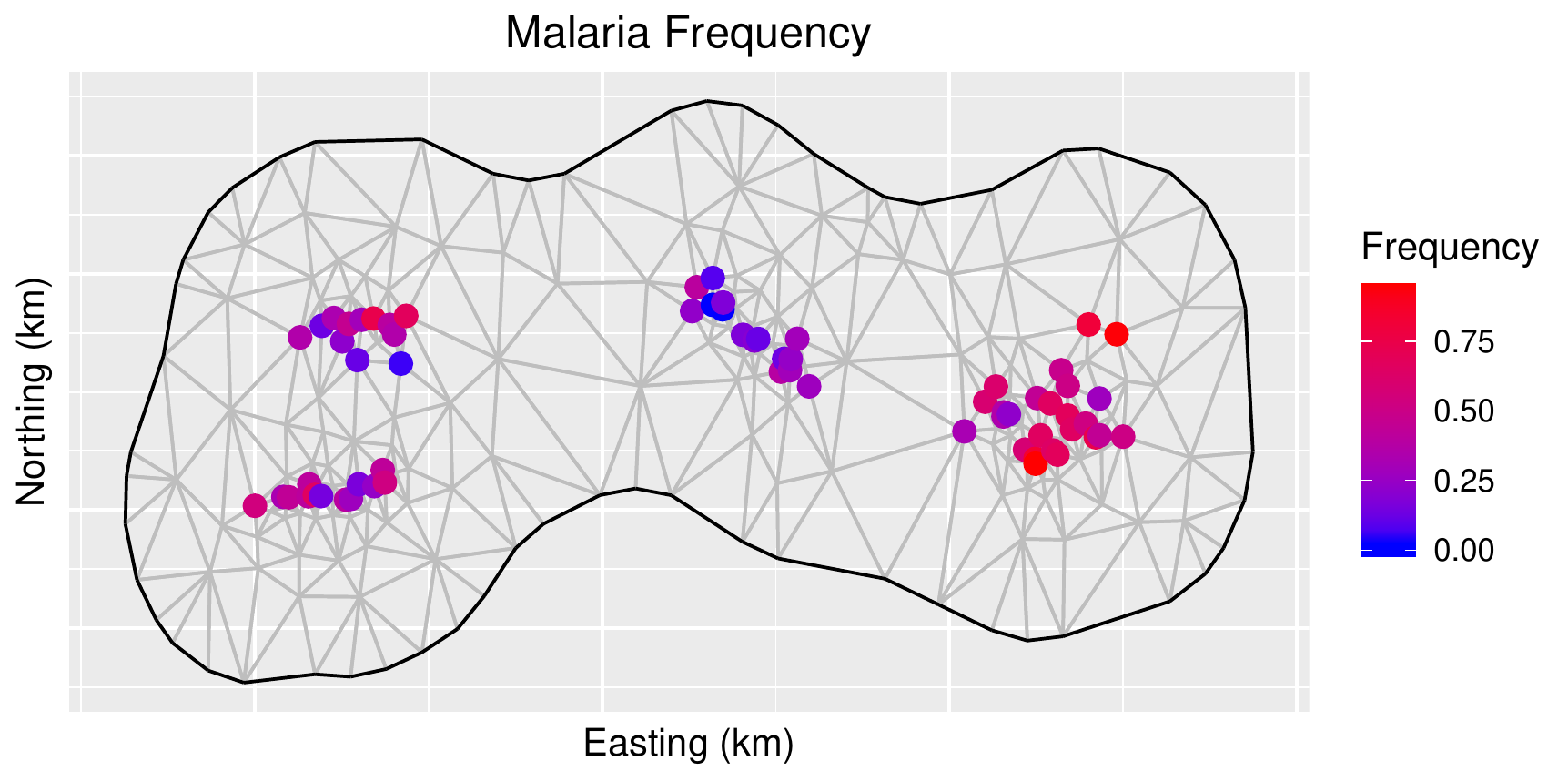}
		\caption{Plot of triangular mesh with $n=288$ nodes and malaria frequency at 65 unique village locations.}
		\label{Malaria_Plot}
	\end{figure}

	We construct a triangular mesh with $n=288$ nodes containing all 65 village locations on a node (see Figure \ref{Malaria_Plot}). We specify a binary probit model with $y^{(i)}_j \sim \text{Binom}\left(1, p_j^{(i)} \right)$, where the probability of malaria presence in the $j^{th}$ child at village location $i$ is linked through the probit function (standard normal CDF) and auxiliary data 
	\bea 
		z^{(i)}_j = {\bs{x}\p_j}^{(i)}\bsbeta + \eta(\bs{u}_i) \label{Malaria_z}
	\eea 
	
	The superscripts for each variable denote the village location $\bs{u}_i$. We fix $\alpha = 2$ leading to covariance matrices for the GRF and LMA priors for $\eta(\bs{u})$ in \eqref{Malaria_z} given by \eqref{cont_Q} and \eqref{K_alpha} respectively. We assume $iid$ normal variance 10 priors for all fixed effects. We follow the data augmentation approach of Albert and Chib \cite{albert1993bayesian} for binary probit GLMs to obtain Gibbs updates for the fixed and random effects. The priors for hyper-parameters $\xi, \kappa, \tau, \lambda$ are all \textit{iid} half normal scale one. Further details pertaining to prior specifications and model fitting are included in Appendix \ref{Appendix_Malaria_Data_Analysis}.
	
	For this dataset, we randomly split the dataset into 10 groups of village locations. All observations associated with a given test set were withheld for validation. Table \ref{Malaria_Table} shows that the LMA model provides the better fit based on BCVS. In this case the LMA also provided a higher ESS. This may be due to the use of a data augmentation approach suggested by of Albert and Chib \cite{albert1993bayesian}. Estimates for all model parameters can be found in Table \ref{Continuous_Analysis_Table}. 
	    
	\begin{table}[H] \centering 
		\begin{tabular}{@{\extracolsep{5pt}} |ccc|} 
			\hline 
			Model & BCVS & ESS \\
			\hline 
			GRF	& 1292.20 & 2.66  \\ 
			LMA	& \textbf{1284.08} &  \textbf{2.90} \\ 
			\hline 
		\end{tabular} 
		\caption{BCVS from ten-fold cross validation and ESS for malaria in the Gambia, Africa. } 
		\label{Malaria_Table} 
	\end{table}

	\subsection{Continuous Space Example: LAGOS } \label{Data_Analysis_LAGOS}
	Lastly, we compare the performance of the LMA for the continuous response LAGOS lake dataset.  The Lake multi-scaled geospatial and temporal database (LAGOS) is a publicly accessible US lake water quality database \cite{soranno2017lagos}. The dataset used in this paper contains records for 5526 unique lakes observed over Iowa, Missouri, and Illinois at locations $\{\bs{u}_i\}_{i=1}^{5526}$. We are interested in modeling the log total phosphorus recorded in each lake. First we reduce the many covariates recorded for each lake in the LAGOS database by performing a step-wise regression assuming uncorrelated residuals. The covariates selected by AIC in this stepwise procedure are: an intercept, lake area (in hectares), max depth of the lake (meters), mean runoff (ground-water discharge into streams), the average annual runoff (in/yr), inter lake water shed (IWS) measurements (the area of land that drains directly into a lake) for urban, agricultural, road density, and total wetland regions. We then used these selected covariates to model log total phosphorus ($y(\bs{u})$) as 
	\bea 
		y(\bs{u}_i) = \bs{x}_i\p \bs{\beta} + \eta(\bs{u}_i) + \eps(\bs{u}_i) 
	\eea    
	
	We fit the LMA model and GRF model by fixing $\alpha=2$. Again this corresponds to Mat\'ern smoothness parameter $\nu=1$. We construct a triangular mesh with $n=671$ nodes according to Section \ref{Subsection_Finite_Element_Approximations_to_Gaussian_Matern_Random_Fields}. The fixed effects $\bs{\beta}$ are assigned a normal prior with variance $10^3$. The nugget term $(\sigma^2)$ and spatial scale parameter $(\kappa)$ are assumed to follow independent half-normal distribution with scale one. The variance term of the GRF $(\xi)$, as well as the variance $(\lambda)$ and scale $(\tau)$ parameters of the LMA are assigned independent half-normal scale one priors. 
	\begin{figure}[H]
		\centering
		\includegraphics[scale=1]{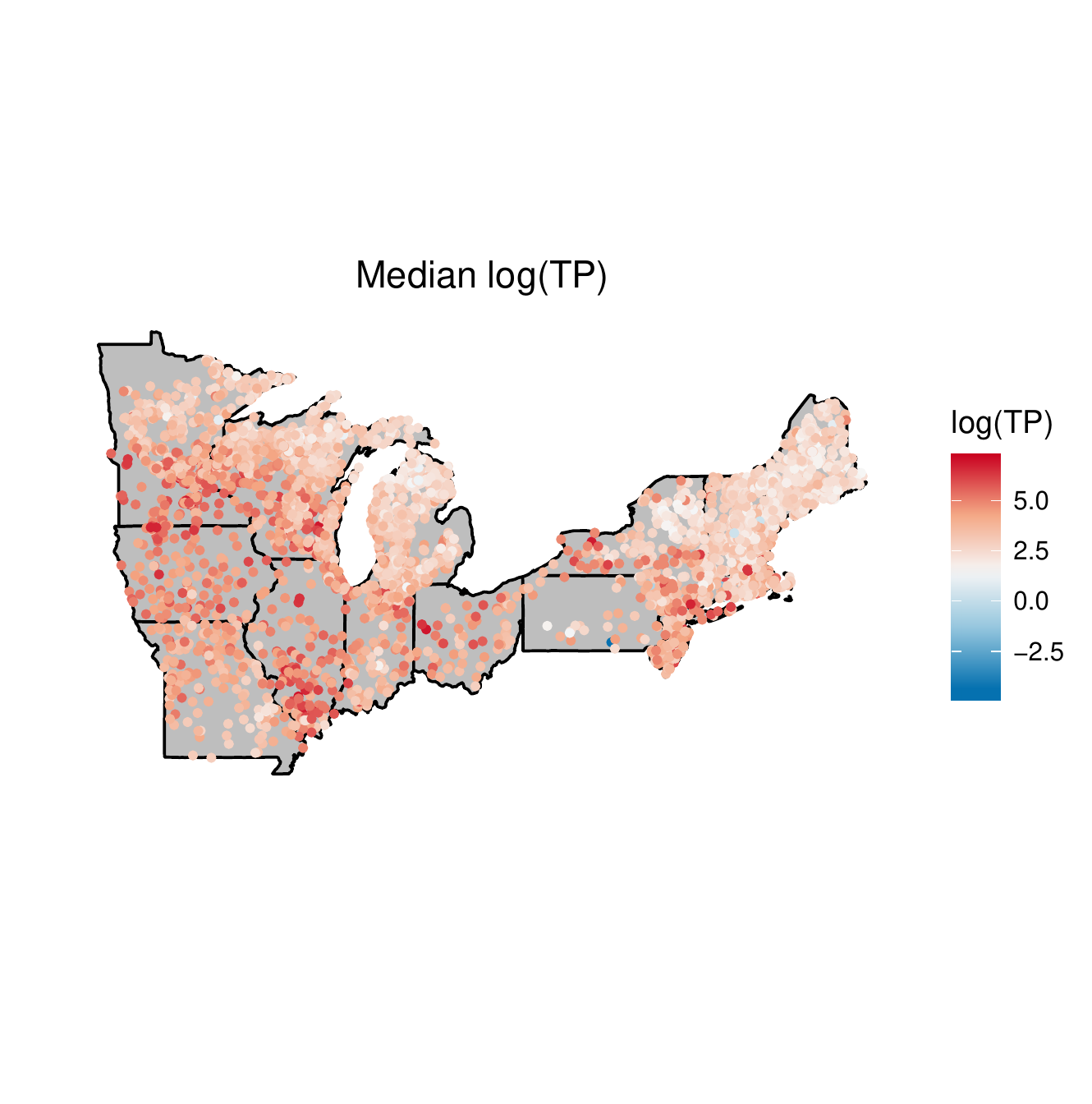}
		\caption{Plot of median log total phosphorus (TP) recorded at 5526 unique lake locations.}
		\label{LAGOS_Plot}
	\end{figure}  
	Table \ref{LAGOS_Table} shows that the LMA model provides a better fit than the GRF in this case. The ESS is better for the LMA as well, further suggesting the LMA model should be favored. As in all other cases, the fixed effect estimates remain roughly similar. All model parameter estimates can be found in Appendix \ref{Appendix_LAGOS} Table \ref{Continuous_Analysis_Table}. Details of the LAGOS data analysis and all full-conditionals are included in Appendix \ref{Appendix_LAGOS}.
	\begin{table}[H] \centering 
		\begin{tabular}{@{\extracolsep{5pt}} |ccc|} 
			\hline 
			Model & BCVS & ESS\\
			\hline 
			GRF	& 5733.309 & 0.35 \\ 
			LMA	& \textbf{5727.993} &  \textbf{0.39} \\ 
			\hline 
		\end{tabular} 
		\caption{BCVS for ten fold cross validation and ESS for the LAGOS dataset.} 
		\label{LAGOS_Table} 
	\end{table}

	\begin{table}[H]\centering  \small 
		\scalebox{0.70}{ \begin{tabular}{@{\extracolsep{5pt}} cccccc} 
				\hline\\[-1.8ex]
				\multicolumn{6}{l}{\textbf{Malaria of the Gambia}} \\
				\hline \\[-1.8ex]
				Predictor & Parameter & GRF Estimate & 95\% CI & LMA Estimate & 95\% CI \\  
				\hline \\[-1.8ex]
				Intercept &$\beta_0$ & 0.123 & (-2.408,  2.677) & -0.975 & (-4.563, 2.554) \\ 
				Age & $\beta_1$ & 0.163  & (0.111,  0.215) &  0.178 & (0.124,  0.232) \\ 
				Bed Net Use & $\beta_2$ & -0.262 & (-0.413, -0.109) & -0.218 & (-0.398, -0.036) \\ 
				Insecticide Use& $\beta_3$ & -0.104 & (-0.270,  0.062) & -0.211 & (-0.417, -0.007) \\ 
				$\log(\text{NDVI})$ & $\beta_4$ & -0.007 & (-0.659, 0.637) &  0.283 & (-0.633,  1.220) \\  
				Health & $\beta_5$ & -0.253 & (-0.397, -0.109) & -0.089 & (-0.297,  0.119) 	\\
				\hline \\[-1.8ex] 
				& $\xi$ & 2.759 & (1.612, 4.203) & NA & NA  \\
				& $\kappa$ & 1.413 & (0.399, 2.433) & 1.192 & (0.518, 1.759)  \\
				& $\tau$ & NA & NA & 13.272 & (12.074, 14.540)\\
				& $\lambda$ & NA & NA & 4.730 & (3.506, 5.824) \\
				\hline \\[-1.8ex]
				\multicolumn{6}{l}{\textbf{LAGOS}} \\
				\hline \\[-1.8ex]
				Predictor & Parameter & GRF Estimate & 95\% CI & LMA Estimate & 95\% CI \\ 
				Intercept & $\beta_0$ & 3.93& (3.773,4.087) &3.925& (3.771,4.081) \\
				Area & $\beta_1$ & 6.33e-05& (5.00e-05,7.71e-05) &6.33e-05& (4.99e-05,7.69e-05)\\
				Max depth & $\beta_2$ & -0.036& (-0.038,-0.034) &-0.036& (-0.038,-0.034)\\
				IWS Urban & $\beta_3$ & 1.296& (1.103,1.489) &1.302& (1.105,1.498)\\
				IWS Ag & $\beta_4$ & 1.594& (1.485,1.701) &1.598& (1.491,1.707)\\
				IWS Wetland & $\beta_5$ & 0.635 & (0.448,0.822) & 0.639 & (0.454,0.826)\\ 
				Road Density & $\beta_6$ & -0.002 & (-0.003,-0.001) &-0.002& (-0.003,-0.001) \\
				Baseflow & $\beta_7$ & -0.012& (-0.014,-0.01) &-0.012& (-0.014,-0.01)\\ 
				Runoff & $\beta_8$ & -0.026& (-0.03,-0.021) &-0.026& (-0.03,-0.021)\\
				\hline \\[-1.8ex] 
				& $\xi$ & 0.383 &( 0.220 , 0.580 ) & NA & NA \\
				& $\sigma^2$ & 0.464 &( 0.444 , 0.485 ) & 0.464 &( 0.444 , 0.485 ) \\ 
				& $\kappa$ & 0.772 &( 0.397 , 1.178 ) & 0.724 &( 0.477 , 0.988 ) \\ 
				& $\tau$ & NA & NA & 1.751 &( 1.322 , 2.263 ) \\
				& $\lambda$ & NA & NA & 0.266 &( 0.184 , 0.38 )
		\end{tabular} }
		\caption{Parameter estimates for continuous space data analysis examples of Sections (\ref{Data_Analysis_Malaria}--\ref{Data_Analysis_LAGOS})}
		\label{Continuous_Analysis_Table} 
	\end{table} 
	
	\section{Discussion} \label{Section_Discussion} 
	
	In this work we proposed a novel discrete space LMA model for irregular lattices and constructed Metropolis Hastings samplers for continuous and discrete space SGLMMs with LMAs. Using the Metropolis Hastings samplers, we provided a broad Bayesian analysis of SGLMMs with LMAs for continuous, binary and Poisson error responses. We found that the LMA offered a better fit than the GRF for datasets which exhibit ``spikes" in the response. Provided our MCMC implementation, we saw that the LMA offers similar computational performance to the GRF.
	
	In this work we restricted our attention to the symmetric LMA. Bolin \cite{bolin2016multivariate} considered more general continuous space LMA which allowed for asymmetric posterior distributions. We note that the asymmetry parameter can be estimated in the MCMC framework, however we elected to compare symmetric models only. We also note that the extension of the asymmetric LMA to discrete space models on irregular graphs could be considered as well. We leave this for future work, but note that the extension should be straightforward.    
	
	The choice of half-normal priors for each of the complexity parameters in the GRF and LMA models, given by $\xi$, $\kappa$, $\tau$, and $\lambda$, were made for the sake of comparison. Fuglstad et al. \cite{fuglstad2015interpretable} proposed the use of a joint prior on the variance $\xi$ and scaling parameter $\kappasq$ in the Gaussian case following the work of Simpson et al. \cite{simpson2017penalising}. We note that their prior choice was motivated by the desire to provide a weakly informative prior and deal with a partial identifiability issue associated with the two parameters. Though identifiability issues with the complexity parameters persist, one does not usually concern themselves with the value of the parameter. We found that independent half-normal priors resulted in the best mixing.   
	
	In summary, we have developed a novel discrete space SGLMM with LMAs. We have proposd the use of Metropolis Hastings samplers to fit the LMA models as a simple extension of the GRF models. Through our extensive data analyses, we have provided evidence of cases in which the LMA model outperforms the traditional GRF SGLMM, while maintaining similar computation efficiency. Due to the comparable computation times and similarity of implementation, we recommend LMA models be considered when modeling correlated error structures.


	\newpage
	\bibliography{CSDS_Bib} 
	
	\newpage
	
	\doublespacing
	\begin{appendices}
		\section{Appendix}
		
		\subsection{CAR Models} \label{Appendix_CAR_Model}
		\noindent \textbf{Definition:} (Conditionally Autoregressive Model) A Conditionally Autoregressive Model (CAR) takes on the form 
		\bea 
			\eta_i | \bs{\eta}_{-i} \sim \mathcal{N}\left(\sum_{\forall C_{ij} \neq 0} C_{ij} \eta_j, M_{ii} \right) \label{CAR_given}
		\eea 
		where \textbf{C} is the spatial dependence matrix with $C_{ii} = 0$, and \textbf{M} is a diagonal matrix with entries $M_{ii}$. The conditional mean of each $\eta_i$ is determined by a weighted sum of neighboring $\eta_j$'s. Note that each marginal variance, $M_{ii}$, varies, so \textbf{M} is often non stationary. 
		
		The CAR model in \eqref{CAR_given} was shown to lead to the full distribution of $\bseta$ by Besag (1974). For positive definite $\bs{Q}^{-1} = (\bs{I}-\bs{C})^{-1}\bs{M}$, \eqref{CAR_given} leads to the full distribution $\bseta \sim N(0,\bs{Q}^{-1})$. Matrices \textbf{M} and \textbf{C} are defined from \textbf{Q} as follows: Write, $\bs{Q} = \bs{D} - \bs{R}$ with 
		\begin{equation*}
			R_{ij} = \begin{cases}
			0,  & \text{if } i = j \\
			-Q_{ij},  & \text{if } i \neq j
			\end{cases}, \quad  \quad
			\bs{D} = \begin{cases}
			Q_{ii},  & \text{if } i = j \\
			0,  & \text{if }  i \neq j 
			\end{cases}.
		\end{equation*}
		This gives $\bs{M} = \bs{D}^{-1}$ and $\bs{C} = \bs{DR}$ in \eqref{CAR_given}. 		
		\subsection{SAR Models} \label{Appendix_SAR}
		Ver Hoef et al. \cite{ver2018relationship} summarized the relationships between SAR and CAR models. CAR and SAR models are widely used in both temporal and spatial statistics due to their intuitive dependence structures. We provide a brief summary of the SAR model for the unfamiliar audience. 
		
		Consider a collection of random variables at \textit{n} spatial locations or graph nodes, $\bs{Y} = (Y_1...,Y_n)$. Let $\bs{\Lambda}$ be a positive diagonal matrix. A SAR model imposes an explicit spatial dependence structure,
		\bea 
			\bs{Y} = \bs{B}\bs{Y} + \bs{\nu}, \quad \bs{\nu} \sim \mathcal{N}\left(\bs{0},\bs{\Lambda} \right) \label{Eqn_SAR_Model}
		\eea 
		where \textbf{B} is a spatial dependency matrix that is not necessarily symmetric. Note that \textbf{B} relates $\bs{Y}$ to itself, and no site can depend on itself, so $B_{ii}$ must be zero for all \textit{i}. Solving for $\bs{Y}$ in \eqref{Eqn_SAR_Model} we have, $(\bs{I}-\bs{B}) \bs{Y} \sim \mathcal{N}(0,\bs{\Lambda})$. The covariance of the SAR model can then be written as $(\bs{I}-\bs{B})^{-1}\bs{\Lambda}[(\bs{I}-\bs{B})\p]^{-1}$, provided $(\bs{I}-\bs{B})$ is invertible. For a thorough comparison of SAR and CAR models see Ver Hoef et al. \cite{ver2018relationship}.  
		
		\subsection{Conditionally Independent Block Proposals}  \label{Appendix_Conditionally_Independent_Block_Proposals} 
		
		Consider $\bs{\eta} \sim N(\bs{0}, \bs{Q}^{-1})$, where $\bs{Q}$ is a GMRF. Define $N(i) := \{j: Q_{ij} \neq 0 \}$. $N(i)$ is the collection of indices \textit{j}, such that \textit{i} and \textit{j} are neighboring points in the spatial lattice. $\eta$ can be expressed as a CAR model with sparse \textbf{Q}
		\bea 
			\eta_i | \bs{\eta}_{-i} = \eta_i | \bs{\eta}_{N(i)} \sim \mathcal{N}\left(\sum_{j \neq i } C_{ij} \eta_j, M_{ii}\right) \label{Eqn_neighbor_cond}
		\eea 
		where the procedure to obtain matrices \textbf{M} and \textbf{C} are described in Appendix \ref{Appendix_CAR_Model}.
		
		To produce one at a time Metropolis Hastings samples, we consider grouping subsets of conditionally independent $\eta_i$'s into blocks. Let $A_k$ be the collection of indices such that, for all $i,j \in A_k$
		\bea 
			\eta_i | \bs{\eta}_{N(i)} \quad \indep \quad \eta_j | \bs{\eta}_{N(j)} \quad \text{ and }  \quad N(i) \cup N(j) \subset A_k^{c} \nt 
		\eea 
		We can now perform one at a time Metropolis Hastings updates for each individual $\eta_i$ within each block $A_k$.

		\subsection{Details of Slovenia Data Analysis} \label{Appendix_Poisson_Data}
		We consider the Poisson SGLMM of form 
		\bea 
			y_i &\sim& Poisson(\mu_i) \nt \\
			\log(\mu_i) &=& \log(o_i) + \bs{x}_i\p \bsbeta + \eta_i + \eps_i \nt 
		\eea 	
		where $o_i$ is an offset for individual $i$.
		
		We use an order $k=1$ differencing matrix for the GRF and LMA model (see equation \eqref{traingle_operator} of Section \ref{Subsection_GRFs_and_LRFs_in_Discrete_Space}). We use the conditional distribution form of the CAR model to perform one at a time block Metropolis Hastings updates following the results of Appendix \ref{Appendix_CAR_Model}--\ref{Appendix_Conditionally_Independent_Block_Proposals}. Matrices \textbf{M} and \textbf{C} of Appendix \ref{Appendix_CAR_Model} are defined for the GRF and LMA models from sparse $\bs{Q}_1$ of equation \eqref{discrete_Q}. Denote ${\eta}_{\mu_i} = \sum_{C_{ij} \neq 0 } C_{ij} \eta_j$. The full-conditionals for $\bseta$ are given by 
		\bea 
			\log\left([ \eta_i | \bseta_{N(i)},y_i,\bsbeta,\eps_i,\bs{\theta} ]\right) &\approx& \log[y_i|\mu_i ] + \log[\eta_i|\bseta_{N(i)},\bs{\theta}] + Const \label{Eqn_Appendix_Poisson_eta} \\
			&\approx& y_i \log(\mu_i) - \mu_i -\frac{(\eta_i - {\eta}_{\mu_i})^2}{2 m_{ii}} + Const 
		\eea 
		In \eqref{Eqn_Appendix_Poisson_eta}, $\bs{\theta} = (\kappasq,\xi)$ for the GRF model and $\bs{\theta} = (\kappasq,\lambda,\bs{S})$ for the LMA. 
		
		A normal prior with variance $\sigma_{\beta}^2 = 10^6$ is assumed for the fixed effects giving log full-conditionals
		\bea 
			\log[\bsbeta|\bsy,\bseta,\bseps] \approx \sum_{i=1}^{194} y_i  \log(\mu_i) - \mu_i - \frac{\bsbeta\p\bsbeta}{2\sigma^2_{\beta}} + Const \nt 
		\eea 
		The prior for the homogeneous spatial random effect is assumed to be \textit{iid} $\mathcal{N}(0,\sigma^2)$, giving log full-conditionals 
		\bea 
			\log[\eps_i | \eta_i, \bsbeta, y_i,\sigma^2] \approx y_i \log(\mu_i) - \mu_i -\frac{\eps_i^2}{2 \sigma^2} + Const \nt 
		\eea 
		An inverse gamma prior with scale and shape one is assumed for $\sigma^2$ giving conjugate full-conditional 
		\bea 
			\sigma^2 \sim InvGamma(98, \frac{||\bseps||^2 }{2} + 1 ) \nt 
		\eea  
		The priors for the variance parameter $(\xi)$ and $(\kappa)$ for the GRF model are assumed to be independent scale one half-normals. The log full-conditionals are 
		\bea 
			\log([\xi | \bs{w},\kappasq]) &\approx& -97\log(\xi^2) - \frac{1}{2\xi^2}\bs{w}\p \bs{LL} \bs{w} - \frac{\xi^2}{2} + Const \nt \\
			\log\left(\left[\kappa| \xi,\bseta \right] \right) &\approx& 2\sum_{i=1}^{194} \log(U_{ii}) - \frac{1}{2\xi^2} \bs{w}\p \bs{LLw} -\frac{\kappa^2}{2} + Const \label{Appendix_Slovenia_GRF_Kappa}
		\eea 
		where $U_{ii}$ in \eqref{Appendix_Slovenia_GRF_Kappa} is the $i^{th}$ diagonal entry of the Cholesky decomposition of $\bs{L}$. 
		The scale parameter, $\lambda$, and $\kappa$ of the LMA have log full-conditionals 
		\bea 
			\log([\lambda|\bs{S}]) &\approx& -194\log(\lambda^2) - \frac{1}{2\lambda^2} \sum_{i=1}^{n} s_{ii} - \frac{ \lambda^2 }{2} + Const \nt \\
			\log\left(\left[\kappa| \bs{S},\bseta \right] \right) &\approx& 2\sum_{i=1}^{194} \log(U_{ii}) - \frac{1}{2} \bs{w}\p \bs{L}\bs{S}^{-1}\bs{Lw} -\frac{\kappa^2}{2} + Const \nt
		\eea 
		\subsection{Details of the Columbus Crime Dataset Analysis} \label{Appendix_Columbus_Crime_Data}
		The model is of form
		\bea 
			\log(y_i) = \bs{x}_i\p\bsbeta + \eta_i + \eps_i \label{Appendix_Columbus_Model}
		\eea 
		where the covariates and response are detailed in Section \ref{Data_Analysis_Columbus}. The fixed effects were assigned a normal prior with variance $10^6$ giving conjugate full-conditionals.
		\bea 
			[\bsbeta|\bsy,\bseta,\sigma^2] &\sim& \mathcal{N}\left( \left( \left(\frac{\sigma^2}{10^6}\right)\textbf{I} + \bsX\p\bsX \right)^{-1} (\bsy - \bseta), \left( \left(\frac{1}{10^6}\right)\textbf{I} + \left(\frac{1}{\sigma^2}\right)\bsX\p\bsX \right)^{-1} \right) \nt 
		\eea  
		The variance of the spatially homogeneous random effect $(\sigma^2)$ is given a half-normal scale one prior. The log full-conditional is 
		\bea 
			\log\left( \left[ \sigma^2 |\bsy,\bseta,\bsbeta\right] \right) &\approx& -\frac{49}{2}\log(\sigma^2) - \frac{||\bs{y}-\bs{X}\bsbeta -\bseta||^2}{2\sigma^2} - \frac{\sigma^2}{2} + Const \nt 
		\eea 
		We use an order $k=1$ differencing matrix to define the covariance structure of the GRF and LMA model (see equation \eqref{traingle_operator} of Section \ref{Subsection_GRFs_and_LRFs_in_Discrete_Space}). The conjugate full-conditionals for the GRF random effects are 
		\bea  
			[\bseta | \bsbeta,\bsy,\kappasq,\xi] &\sim& \mathcal{N}\left( \left( \left(\frac{\sigma^{2}}{\xi^2} \right) \bs{LL} + \textbf{I} \right)^{-1} (\bsy - \bsX\bsbeta), \left( \left(\frac{1}{\xi^2} \right) \bs{LL} + \left(\frac{1}{\sigma^2}\right) \textbf{I} \right)^{-1} \right) \nt
		\eea  
		The scale parameter for the GRF $(\xi)$ is given a half-normal scale 10 prior while $\kappa^2$ is given an independent half-normal scale one prior leading to log full-conditionals 
		\bea   
			\log\left( \left[\xi^2 | \bseta,\kappasq  \right] \right) &\approx& -\frac{49}{2}\log(\xi^2) - \frac{1}{2\xi^2} \bseta\p \bs{LL} \bseta - \frac{\xi^4}{20} + Const \nt \\
			\log\left( \left[\kappasq | \bseta,\xi \right] \right) &\approx& 2\sum_{i=1}^{49} \log(U_{ii}) - \frac{1}{2\xi^2} \bseta\p \bs{LL} \bseta - \frac{\kappa^4}{2} + Const \nt  
		\eea 
		where $U_{ii}$ is the Cholesky decomposition of $\bs{L} = \triangle^{(1)}$. The conjugate full-conditionals for the LMA random effects are 
		\bea 
			\left[\bseta | \bsbeta,\bsy,\kappasq,\bs{S}\right] &\sim& \mathcal{N}\left( \left( \sigma^{2} \bs{L}\bs{S}^{-1}\bs{L} + \textbf{I} \right)^{-1} (\bsy - \bsX\bsbeta), \left(\bs{L}\bs{S}^{-1}\bs{L} + \left(\frac{1}{\sigma^{2}}\right)\textbf{I} \right)^{-1} \right) \nt
		\eea 
		The scale parameter $(\lambda)$ for the LMA is given a half-normal scale 10 prior and $\kappasq$ is given an independent scale one half-normal prior. The log full-conditionals for LMA model parameters are  
		\bea
			\log( [\lambdasq | \bs{S}  ] ) &\approx& -49\log\left(\lambdasq\right) - \frac{1}{2\lambdasq} \sum_{i=1}^{49} S_{ii} - \frac{\lambda^4}{20} + Const \nt \\
			\log([\kappasq | \bseta,\bs{S}]) &\approx& 2\sum_{i=1}^{49} \log(U_{ii}) - \frac{1}{2} \bseta\p \bs{L}\bs{S}^{-1}\bs{L} \bseta - \frac{\kappa^4}{2} + Const \nt 
		\eea 
		We observed spatial confounding among the random effects and the intercept. This is not uncommon, however to assess convergence we analyzed the trace plots of $\beta_0 \bs{1} + \bseta$.

		\subsection{Details of Malaria Data Analyses} \label{Appendix_Malaria_Data_Analysis}
		We follow the auxiliary data approach of Albert and Chib \cite{albert1993bayesian}. Let $\Phi(\cdot)$ denote the standard normal CDF. Consider the continuous space binary response model $y^{(i)}_j \sim Bernoulli(p^{(i)}_j)$ where $p_j^{(i)}$ is the probability that the $j^{th}$ child in the $i^{th}$ village has malaria. We model $p_j^{(i)}$ through the probit link function by introducing auxiliary data $z^{(i)}_j $ as follows  
		\bea 
			p^{(i)}_j = \Phi(z^{(i)}_j), & & z^{(i)}_j = {\bs{x}\p}^{(i)}_j \bsbeta + \eta(\bs{u}_i). \nt  
		\eea 
		The covariates are as described in Section \ref{Data_Analysis_Malaria}.	Define $A_{ij} = \phi_j(\bs{u}_i)$, where $\{\phi_l(\bs{u})\}_{l=1}^{n}$ are the basis functions corresponding to the triangular mesh with $n=288$ mesh nodes formed in Section \ref{Data_Analysis_Malaria}. Define the 2035 by 65 matrix \textbf{B} such that $n_i$ entries of column $\bs{b}_i$ corresponding to responses $y^{(i)}_j$ are 1, and the remaining entries are 0. The auxiliary variables can be equivalently expressed in matrix form as
		\bea 
			\bs{z} = \bs{X} \bsbeta + \bs{BAw}. 
		\eea 
		
		The priors for $\bs{w}$ are constructed following Sections (\ref{Subsection_Finite_Element_Approximations_to_Gaussian_Matern_Random_Fields}-\ref{Subsection_Finite_Element_Approximations_to_Matern_LMAs}) with $\alpha = 2$. We place a normal prior on the fixed effects with variance 100. In both models the spatial scale parameter $\kappa$ was assigned a half-normal prior with scale one. Additionally, in the Gaussian model, the scale parameter $(\xi)$ was assigned a half-normal scale one prior as well. We used joint MH proposals for $\kappa$ and $\xi$ in the Gaussian model. For the LMA, shape parameter $(\tau)$ and scale parameter $(\lambda)$ were jointly proposed with independent scale one half-normal priors.

		\noindent \textbf{Full-Conditionals}\\
		For both models, we have conjugate truncated normal ($TN_{(a,b)}$) updates for the auxiliary variables,
		\[ \left[z^{(i)}_j| y^{(i)}_j,w_i, \bsbeta \right] \sim \begin{cases} 
			TN_{(0,\infty)}, & z^{(i)}_j > 0 \\
			TN_{(-\infty,0)}, & z^{(i)}_j < 0
			\end{cases}
		\]
		and conjugate normal updates for the fixed effects,
		\bea 
			[\bs{\beta}|\bs{w},\bs{y},\bs{z}] \sim \mathcal{N}\left( \left[\bs{X}\p\bs{X} + \left(\frac{1}{100}\right)\textbf{I} \right]^{-1}(\bs{y}-\bs{BA}\bs{w}), \left[\bs{X}\p\bs{X} + \left(\frac{1}{100}\right)\textbf{I} \right]^{-1} \right). \nt 
		\eea 
		The weights of the basis expansion for the GRF are given by 
		\bea 
			[\bs{w}|\bsbeta,\bs{y},\bs{z},\xi,\kappa] \sim \mathcal{N}\left( \left[ \left(\frac{1}{\xi^2}\right) \bs{LL} + \bs{A}\p \bs{B}\p \bs{BA} \right]^{-1}(\bs{y}-\bs{X}\bsbeta),  \left[\left(\frac{1}{\xi^2}\right) \bs{LL} + \bs{A}\p \bs{B}\p \bs{BA} \right]^{-1} \right). \nt
		\eea 
		The full-conditional for the spatial scale $(\kappa)$ and variance parameter $(\xi)$ are given by 
		\bea 
			\log([\kappa,\xi|\bs{w}]) \approx -\left(\frac{1}{2\xi^2}\right)\bs{w}\p \bs{LL} \bs{w} - \left(\frac{n}{2}\right)\log(\xi^2) + 2\sum_{i=1}^{n}\log(U_{ii}) - \frac{\kappasq}{2} - \frac{\xi^2}{2} + Const \label{Appendix_kap_xi_conditional}
		\eea 
		where $U_{ii}$ denotes the $i^{th}$ diagonal entry of the Cholesky of $\bs{L}=\kappasq \bs{C} + \bs{G}$.
		The weights of the basis expansion for the LMA are given by 
		\bea 
			[\bs{w}|\bsbeta,\bs{y},\bs{z},\bs{\Gamma},\kappa] \sim \mathcal{N}\left( \left[\bs{L}\bs{\Gamma}^{-1}\bs{L} + \bs{A}\p \bs{B}\p \bs{BA} \right]^{-1}(\bs{y}-\bs{X}\bsbeta),  \left[\bs{L}\bs{\Gamma}^{-1}\bs{L} + \bs{A}\p \bs{B}\p \bs{BA} \right]^{-1} \right). \nt
		\eea 
		The spatial scale $(\kappa)$, shape $(\tau)$, and variance parameter $(\lambda)$, have log full-conditionals 
		\bea 
			\log\left([\kappa,\tau,\lambda|\bs{\Gamma},\bs{w}]\right) &\approx& \sum_{i=1}^{n} \left( (\tau C_{ii})\left(\log(\Gamma_i)-\log\left(\lambdasq\right)\right) - \log(\Gamma(\tau C_{ii})) - \frac{\Gamma_{ii}}{\lambdasq} + 2\log(U_{ii}) \right) \nt \\
			& & - \frac{1}{2}\bs{w}\bs{L}\bs{\Gamma}^{-1}\bs{L} \bs{w} - \frac{\lambdasq}{2} - \frac{\kappasq}{2} - \frac{\tau}{2} + Const \nt  
		\eea 
		
		\subsection{Details of LAGOS Analysis}  \label{Appendix_LAGOS} 
		We fit a continuous response point referenced model with 5526 unique lake locations denoted $\{\bs{u}_i\}_{i=1}^{5526}$. We form a mesh with $n = 671$ nodes. The model considered is of the form 
		\bea 
			y(\bs{u}_i) = \bs{x}_i\p \bsbeta + \eta(\bs{u}_i) + \eps(\bs{u}_i).
		\eea 
		Define the 5526 by 671 projection matrix $(\bs{A})$ with entries $A_{ij} = \phi_j(\bs{u}_i)$, where $\{\phi_l(\bs{u})\}_{l=1}^{671}$ are the basis functions corresponding to the mesh formed in Section \ref{Data_Analysis_Malaria}. Using the resulting basis expansion of $\eta(\bs{u})$ (see \eqref{basis_expansion} Section \ref{Subsection_Finite_Element_Approximations_to_Gaussian_Matern_Random_Fields}) and assuming $\eps(\bs{u}_i) \stackrel{iid}{\sim} \mathcal{N}(0,\sigma^2)$, we can re-write the discretized likelihood as follows 
		\bea 
			[\bs{y}|\bsbeta,\bs{w},\sigma^2] \sim \mathcal{N}\left( \bs{X}\bsbeta + \bs{A}\bs{w}, \sigma^2 \bs{I} \right).   
		\eea 
		The priors for $\bs{w}$ are constructed following Sections (\ref{Subsection_Finite_Element_Approximations_to_Gaussian_Matern_Random_Fields}-\ref{Subsection_Finite_Element_Approximations_to_Matern_LMAs}) with $\alpha = 2$. We have assumed half-normal scale one priors for $\sigma^2, \kappa, \xi, \tau$, and $\lambda$. \\
		\textbf{Full-Conditionals}	\\
		The fixed effects for both models have conjugate full-conditionals 
		\bea 
			[\bsbeta|\bsy,\sigma^2,\bs{w}] \sim \mathcal{N}\left( \left[ \left(\frac{\sigma^2}{1000}\right) \bs{I} + \bs{X}\p\bs{X} \right]^{-1}\left[ \bsy - \bsX\bsbeta\right], \left[ \left(\frac{1}{1000}\right) \bs{I} + \left(\frac{1}{\sigma^2}\right)\bs{X}\p\bs{X} \right]^{-1} \right). \nt
		\eea 
		The conjugate full-conditionals for the weights of the GRF are
		\bea 
			\left[\bs{w}|\bsy,\sigma^2,\bsbeta,\xi \right] \sim \mathcal{N}\left( \left[ \left(\frac{\sigma^2}{\xi^2}\right) \bs{L}\bs{C}^{-1}\bs{L} + \bs{A}\p \bs{A} \right]^{-1} \left[ \bsy - \bsX\bsbeta\right], \left[ \left(\frac{1}{\xi^2}\right) \bs{L}\bs{C}^{-1}\bs{L} + \left(\frac{1}{\sigma^2}\right) \bs{A}\p \bs{A} \right]^{-1} \right). \nt
		\eea 
		The log full-conditionals for $\kappa$ and $\xi$ of the GRF model are as seen in equation \eqref{Appendix_kap_xi_conditional} of Appendix \ref{Appendix_Malaria_Data_Analysis}. The conjugate full-conditionals for the weights of the LMA are 
		\bea 
			\left[\bs{w}|\bsy,\sigma^2,\bsbeta,\bs{\Gamma} \right] \sim \mathcal{N}\left( \left[ \sigma^2 \bs{L}\bs{\Gamma}^{-1}\bs{L} + \bs{A}\p \bs{A} \right]^{-1} \left[ \bsy - \bs{A}\bs{w} \right], \left[  \bs{L}\bs{\Gamma}^{-1}\bs{L} + \left(\frac{1}{\sigma^2}\right) \bs{A}\p \bs{A} \right]^{-1} \right). \nt
		\eea 
		The log full-conditionals for the parameters $\kappa, \tau$ and $\lambda$ are as seen in \eqref{Appendix_kap_xi_conditional} of Appendix \ref{Appendix_Malaria_Data_Analysis}. For model fitting we used normal proposals for all parameters. $\tau$ and $\lambda$ were jointly proposed for the LMA model, while $\kappa$ and $\xi$ were jointly proposed for the GRF.

	\end{appendices}

\end{document}